\documentclass[reprint,onecolumn,amsmath,amssymb,aps,superscriptaddress]{revtex4-2}

\usepackage{graphicx}
\usepackage{dcolumn}
\usepackage{bm}
\usepackage{graphicx}
\usepackage{amssymb}
\usepackage{epstopdf, epsfig}
\usepackage{yfonts}
\usepackage[colorlinks=true, linkcolor=blue, citecolor=blue, urlcolor=blue]{hyperref}
\usepackage{fontenc}
\usepackage{upgreek}
\usepackage{mathrsfs}
\usepackage{booktabs}
\usepackage{pifont}
\usepackage{amsmath}
\usepackage{bm}
\usepackage{wasysym}
\usepackage{caption}
\usepackage{subcaption}
\usepackage{euscript}
\usepackage{natbib}
\usepackage{lipsum}
\usepackage{xcolor}
\usepackage{multirow}
\usepackage{cases}

\usepackage{orcidlink}

\usepackage{subcaption}
\usepackage{ragged2e}
\DeclareCaptionJustification{justified}{\justifying}
\newcommand{\x}[1]{\textcolor{black}{#1}}
\newcommand{\xx}[1]{\textcolor{black}{#1}}

\newcommand{\jfm}[1]{\textcolor{blue}{#1}}

\begin{document}

\title{Evolution of extreme nonlinear wave fields over strongly reflective plane beaches}

\author{Jie Zhang\,\orcidlink{0000-0003-0794-2335}}
\email{jie.zhang@hrbeu.edu.cn}
\affiliation{College of Shipbuilding Engineering, Harbin Engineering University, Harbin 150001, China}
\affiliation{National Key Laboratory of Hydrodynamics, Harbin Engineering University, Harbin 150001, China}
\affiliation{Qingdao Innovation and Development Center of Harbin Engineering University, Qingdao 266000, China}

\author{Michel Benoit\,\orcidlink{0000-0003-4195-2983}}
\email{\jfm{michel.benoit@edf.fr}}
\affiliation{EDF R\&D, Laboratoire National d’Hydraulique et Environnement (LNHE), Chatou 78400, France}
\affiliation{LHSV, ENPC, Institut Polytechnique de Paris, EDF R\&D, Chatou 78400, France}

\author{Saulo Mendes\,\orcidlink{0000-0003-2395-781X}}
\email{saulo.mendes@ntu.edu.sg}
\affiliation{School of Civil \& Environmental Engineering, Nanyang Technological University, Singapore}

\begin{abstract}
The description of complex wave processes, in addition to the shoaling problem, is often cumbersome even for the evolution of regular waves. For reflection under the regime of wave breaking, the surf similarity is generally accepted as the leading parameter controlling the reflection rates and types of breakers. Although little is known about the effect of reflection rates on the formation of extreme nonlinear waves, some debate has arisen regarding whether high reflection rates enhance the nonlinearity at the tail of the wave height distribution. In this work, we provide theoretical evidence that at very steep beaches of smooth composition, the reflection rate near unity will tend to stabilize the excess kurtosis otherwise generated by shoaling and controlled in magnitude by the bottom slope magnitude. We further verified this result through fully nonlinear numerical simulations, reaching a good agreement.
\end{abstract}


\maketitle

\section{INTRODUCTION}\label{sec:intro}

The study of nonlinear waves presents applications in many fields and has become one of the most eclectic areas of the physical sciences. From geophysics and astrophysics to oceanography, meteorology, and even engineering or optics, the behaviour of these waves is, on the one hand, unique in their medium of propagation, but on the other hand, they share common properties \citep{Hasegawa1974,Lerche1982,Godano1999,Whitham2011,Ogilvie2016,Dudley2019,Kulikovskii2021} and mathematical descriptions \citep{Kadomtsev1971,Zakharov1997}. Fundamentally, nonlinear processes are enhanced when out-of-equilibrium dynamics takes place, often by disturbances in material or structural composition (e.g. heterogeneous media). For water waves, non-homogeneous bathymetry and the elasticity thereof can turn the propagation of nonlinear waves into a quagmire \citep{Liu2007,Tsai2013,Mei2013,Reniers2016,Kirby2017,Kirby2019}. Therefore, modeling realistic conditions often requires simplification of the effects that are not of leading order while attaining crucial components. Ultimately, no theoretical hydrodynamics model may contain all known effects, such as those due to non-homogeneity or non-stationarity, elasticity, viscosity, plasticity, reflection, diffraction, refraction, shoaling, or breaking.

The deterministic treatment of fundamental and engineering properties of water waves is expected to be less convoluted than the stochastic treatment thereof. When adding far too many wave processes into the same picture, the deterministic approach is considerably simpler than the stochastic one. This is in part because the wave properties investigated in the regular wave field case only materializes in the bulk of the wave height distribution, so much so that the physics of the modulation of absolute variables (such as wave height or wavelength) often has no impact on the tail of the height distribution \citep{Stansell2004,Christou2014,Srokosz2018}. For instance, wave breaking for extreme waves has been found to outdo the threshold expected for regular waves \citep{Babanin2011}. While sediment transport rates are prone to enhancement due to nonlinear waves \citep{Damgaard2005}, extreme waves at the tail are significantly more asymmetrical \citep{Christou2014,Mendes2021a} and thus can induce faster transport. Recently, \citet{Yuchen2023} has investigated the effect of reflection due to a beach at several angles and composed of different materials. While reflection is known to enhance the significant wave height, especially against a wall, \citet{Yuchen2023} observed fewer extreme nonlinear wave heights. Typically, wave crests have a stronger response to nonlinearities, and the deviation from the Rayleigh distribution \citep{Higgins1952} is more markedly seen for wave crests than wave heights. The latter may explain that \citet{Babanin2020} has found a two-fold increase in crest height wave statistics against a vertical wall. Additional observations have found that rogue waves formed over steep slopes with high reflection tend to be much stronger and more dangerous than those formed over mild beach slopes \citep{Didenkulova2023}. In other words, understanding the physics of nonlinear normalized heights (or crest heights) is not trivial, even if one knows the behavior of the wave height itself.

The behavior and stochastic outcomes of irregular waves traveling past a shoal or a breakwater have been thoroughly experimented and discussed for the ideal case of a smooth bathymetry \citep{Zhang2019,Bolles2019,Mendes2022,Chabchoub2023}. Recently, a new experimental direction has been pursued, and investigations on the coupled effect of waves and currents over a shoal have been examined \citep{Benoit2023}. Indeed, \citet{Yuchen2023} has provided seminal evidence that the smoothness of the bathymetry affects the effect of reflection on extreme wave statistics: on a smooth beach the vertical wall slightly enhances the chances of finding rogue wave heights, albeit smaller reflection rates due to angles at $8^{\circ}$ feature the highest amplification. On a grass beach, \citet{Yuchen2023} found the same trend (reflection increases the likelihood of extreme waves to a certain degree) that saturates at steep angles, and ultimately, the reflection due to a vertical wall reduces extreme nonlinear waves. These findings generalize the main understanding of \citet{Babanin2020}, such that depending on the reflection rate and smoothness of the beach, one may or may not find an increase of the already existing excess kurtosis due to shoaling. 

These latter two investigations focus on the reflection rate affecting the statistics of waves set before the wall/shoal. In this work, we are interested in finding how reflection affects the extreme wave statistics on the plateau of a \xx{submerged} breakwater in shallow water, \textit{in lieu} of a plane beach. In this regard, the literature is rather limited. The remainder of the paper is organized as follows: in \jfm{section}~\ref{sec:ell_on_PDF}, we present a theoretical model for the effect of shoaling length (i.e. how large is the shoal compared to the characteristic wavelength) building on previous works of the authors. The model is then validated for cases with varying slope lengths and a constant slope using fully nonlinear numerical simulations. In \jfm{section}~\ref{sec:validation}, we show the numerical evidence that a high reflection rate may limit the enhancement of the kurtosis. Then we extend our formula to plane beaches with high reflection rates and provide an appropriate explanation for the saturation of the excess kurtosis. Conclusions are given in \jfm{section}~\ref{sec:conclusion}.

\section{Effect of shoaling length on post-shoal wave statistics}\label{sec:ell_on_PDF}

\subsection{Theoretical developments}\label{sec:ell0}

In \citet{Zhang2024} we have shown numerically that the effect of $\ell := L/\lambda$ (shoaling length \x{$L$ normalized by the characteristic wavelength $\lambda$}) at fixed steep slopes is insignificant at best. In this work, before we move to studying the effect of slope of strongly reflective beaches, we study \xx{from an analytical perspective} how the ineligibility of the $\ell$ effect holds for different slope magnitudes. To examine the effect of the shoaling length on wave statistics, we consider the free surface elevation (FSE) $\eta$ and velocity potential $\Phi$ of the second-order Stokes solution in a 2D Cartesian coordinate system $(x,z)$ with its $z$-axis origin located at the still water level (SWL) and pointing upward \citep{Dingemans1997,Zhang2023}:
\begin{eqnarray}
 \eta (x,t) = a\left[\cos{\phi} +    \left( \frac{ \pi\varepsilon }{ 4  } \right) \sqrt{\tilde{\chi}_{1}} \cos{(2\phi)}\right]  \quad , 
\end{eqnarray}
\begin{eqnarray}
 \Phi (x,z,t) = \frac{a\omega }{k} \left[  \frac{\cosh{\varphi} }{\sinh{ \mu }} \sin{\phi} + \left(  \frac{3 \pi\varepsilon}{8}  \right) \frac{\cosh{(2\varphi)} }{\sinh^{4}{\mu}} \sin{(2\phi)}  \right] \quad , 
\end{eqnarray}
where $\varphi \equiv k (z+h)$, $\phi \equiv kx - \omega t$, and $a$, $k$, $\omega$ denote the amplitude, wavenumber and angular frequency of the regular wave, $h$ denotes the dimensional water depth, $\mu\equiv kh$ the relative water depth, and $\varepsilon \equiv 2a/\lambda$ the steepness. The super-harmonic coefficient is:
\begin{equation}
\Tilde{\chi}_{1} = \frac{(3 - \sigma^2)^2} {\sigma^6}  \,\, , \,\, \sigma \equiv \tanh\mu\;,
\label{eq:Rayexc_para}
\end{equation}

In the present configuration, the FSE variance of second-order in steepness \citep{Mendes2022} is approximated as a spatial average over the slope length $L$:
\begin{eqnarray}
\langle \eta^{2} \rangle \approx  \frac{1}{L} \int_{0}^{L}  \eta^{2}(x) \, \textrm{d}x 
= \frac{a^2}{L} \int_{0}^{L} \left[ \cos{\phi} +    \left( \frac{ \pi \varepsilon  }{ 4  } \right) \sqrt{\tilde{\chi}_{1}} \cos{(2\phi)} \right]^2  \, \textrm{d}x  \quad  ,
\label{eq:variance_app}
\end{eqnarray}
and we calculate the total energy $\mathscr{E}$ via an average over $L$ as: 
\begin{eqnarray}
 \mathscr{E}  &\equiv & \mathscr{E}_{p} + \mathscr{E}_{k}  = \frac{1}{2L} \int_{0}^{L}  \eta^{2} (x) \, \textrm{d}x  + \frac{1}{2gL } \int_{0}^{L} \int_{-h(x)}^{0} \left[ \left( \frac{\partial \Phi}{\partial x} \right)^{2} + \left( \frac{\partial \Phi}{\partial z} \right)^{2} \right] \, \textrm{d}z \, \textrm{d}x \,\, ,
 \label{eq:energyX0}
\end{eqnarray}
Note that the definitions in eqs.~\eqref{eq:variance_app} and \eqref{eq:energyX0} are different from those given in eq.~(2.4) of \citet{Mendes2022}, the interval of integration is changed from one wavelength to the length of the shoal $L$ here, to formulate the effect of $\ell$ explicitly.

For the potential energy, its definition turns out to be half of the approximation used for the variance in eq.~\eqref{eq:variance_app}. We stress that the variance of FSE being twice the averaged potential wave energy is not a universal result, but a consequence of the above choice to evaluate the variance in this study. Thus we have:
\begin{eqnarray}
\langle \eta^{2} \rangle = 2\mathscr{E}_{p} &=&  a^{2} \left [  \mathcal{I}_{1} + 2\mathcal{I}_{12} \left(  \frac{\pi \varepsilon}{4} \right) \sqrt{ \Tilde{\chi}_{1} }  + \mathcal{I}_{2} \left(  \frac{\pi \varepsilon}{4} \right)^{2}  \Tilde{\chi}_{1} \right ]
\label{eq:energyVAR}
\\
\nonumber
&=&  \frac{a^{2}}{2} \Bigg[  1 + \left(  \frac{\pi \varepsilon}{4} \right)^{2}  \Tilde{\chi}_{1} + \mathcal{H}_{11\ell}
+ \mathcal{H}_{12\ell}  \left(  \frac{\pi \varepsilon}{4} \right)^{2}  \Tilde{\chi}_{1} + \mathcal{H}_{13\ell}  \frac{\pi \varepsilon}{4} \sqrt{ \Tilde{\chi}_{1} } \Bigg] \quad ,
\end{eqnarray}
where the terms $\mathcal{I}_{1}$, $\mathcal{I}_{2}$, $\mathcal{I}_{12}$, and $\mathcal{H}_{11\ell}$, $\mathcal{H}_{12\ell}$, $\mathcal{H}_{13\ell}$ are the auxiliary coefficients (detailed in eq.~\eqref{eq:setint} and eq.~\eqref{eq:H13l} in \jfm{appendix}~\ref{append:auxilliary_coef}) which all end up in corrections due to $\ell \equiv L/\lambda$.

The kinetic energy $\mathscr{E}_{k}$ is heavier in algebra than $\mathscr{E}_{p}$. To accurately keep track of all terms, we focus on the trigonometric terms, rewriting $\mathscr{E}_{k}$ in eq.~\eqref{eq:energyX0}:
\begin{eqnarray}
\hspace{-1.4cm}
\mathscr{E}_{k} &=&   \frac{(a\omega)^{2}}{2gL }  \int_{0}^{L}     \textrm{d}x  \left\{  \mathcal{J}_{1} \frac{  \cos^{2}{\phi} }{ \sinh^{2}{ \mu } }   + \mathcal{J}_{12}   \frac{3\pi \varepsilon}{2} \frac{ \cos{\phi} \cos{(2\phi)} }{\sinh^{5}{ \mu }} + \mathcal{J}_{2}  \left( \frac{3\pi \varepsilon}{4} \right)^{2} \frac{  \cos^{2}{(2\phi)} }{\sinh^{8}{ \mu }}  \right\} 
\\
\nonumber
&+&  \frac{(a\omega)^{2}}{2gL }  \int_{0}^{L}     \textrm{d}x  \left\{  \mathcal{J}_{3} \frac{  \sin^{2}{\phi} }{ \sinh^{2}{ \mu } }   + \mathcal{J}_{34}   \frac{3\pi \varepsilon}{2} \frac{ \sin{\phi} \sin{(2\phi)} }{\sinh^{5}{ \mu }} + \mathcal{J}_{4}  \left( \frac{3\pi \varepsilon}{4} \right)^{2} \frac{  \sin^{2}{(2\phi)} }{\sinh^{8}{ \mu }}  \right\} , 
\end{eqnarray}
with the set of auxiliary integrals $\mathcal{J}_{1}$, $\mathcal{J}_{2}$, $\mathcal{J}_{12}$, $\mathcal{J}_{3}$, $\mathcal{J}_{4}$, $\mathcal{J}_{34}$ again provided in eq.~\eqref{eq:setint2} in \jfm{appendix}~\ref{append:auxilliary_coef}.

To ease the computation, several steps of algebraic manipulation and leading-order approximations are needed. For instance, we must consider the dispersion relation $\omega^{2} = gk \tanh{ \mu }$, leading to $\big[ (a\omega)^{2}/(2g) \big] (2k\sinh^{2}{ \mu })^{-1} = (a^{2}/4)\big[ \tanh{ \mu } / \sinh^{2}{ \mu } \big] \equiv a^2/(2\sinh{(2 \mu )})$. Moreover, once the identity $\big[ 2 / \sinh{(2 \mu )}\big](\sinh{(4 \mu )} / 4 ) = \cosh{(2 \mu )}$ is realized together with the approximation $(9 \cdot 3/2)\sinh^{-6}{ \mu } \approx \chi_{1}$ at the narrow range of the peak in amplification $ \mu  \sim 1/2$, resorting to the set of integrals from eq.~(\ref{eq:setint}) we finally achieve a full expression for the kinetic energy:
\begin{eqnarray}
\mathscr{E}_{k} =  \frac{a^{2}}{4} \Big[  1   + \left( \frac{\pi \varepsilon}{4} \right)^{2}   \chi_{1}  + \mathcal{H}_{11\ell}  + \mathcal{H}_{12\ell} \left( \frac{\pi \varepsilon}{4} \right)^{2}   \chi_{1}  + \mathcal{H}_{13\ell}\frac{\pi \varepsilon}{4} \sqrt{ \Tilde{\chi}_{1} } f_3 \Big] \,\,\, , \,\,\,  \chi_{1} = \frac{9\, \textrm{cosh}(2\mu) }{\textrm{sinh}^{6} \mu} \,\, ,
\label{eq:A_kine_energy}
\end{eqnarray}
where $\chi_1$ the kinetic energy residue counterpart to the super-harmonic, and $f_{3}$ reads:
\begin{eqnarray}
\nonumber
 f_{3}  &=&  \frac{    3\sinh{ \mu } +  \sinh{(3 \mu )} }{\sinh{(2 \mu )} \sinh^{3}{ \mu } \sqrt{ \Tilde{\chi}_{1} }} + \frac{16}{ \sinh{(2 \mu )} \sqrt{ \Tilde{\chi}_{1} } } \frac{  \sin^{3}{(2\pi \ell)}}{3\sin{(2\pi \ell)} +  \sin{(6\pi \ell)}}  
\\
\nonumber
&+& \left[ \frac{2  \mu  }{\sinh{(2 \mu )}} - 1  \right] \frac{3}{\pi \varepsilon \sqrt{ \Tilde{\chi}_{1} }}  \frac{\sin{(4\pi \ell)} }{  3\sin{(2\pi \ell)} +  \sin{(6\pi \ell)} } \left[ 1 +  \cos(4\pi\ell) \left( \frac{\pi \varepsilon}{4} \right)^{2}   \tilde\chi_{1}  \right]
\\
&-& \frac{2  \mu  }{\sinh{(2 \mu )}} \frac{\sin{(8\pi \ell)} }{3\sin{(2\pi \ell)} +  \sin{(6\pi \ell)}} \frac{\pi \varepsilon}{4 \sqrt{ \Tilde{\chi}_{1} }}  \frac{  27  }{16 \sinh^{6}{ \mu }}.
\end{eqnarray}

Wherefore, summing up the results of eqs.~\eqref{eq:energyVAR} and \eqref{eq:A_kine_energy}, the total energy $\mathscr{E} = \mathscr{E}_{k}+ \mathscr{E}_{p}$ becomes:
\begin{figure*}
\minipage{0.4\textwidth}
\hspace{0.7cm}
    \includegraphics[scale=0.6]{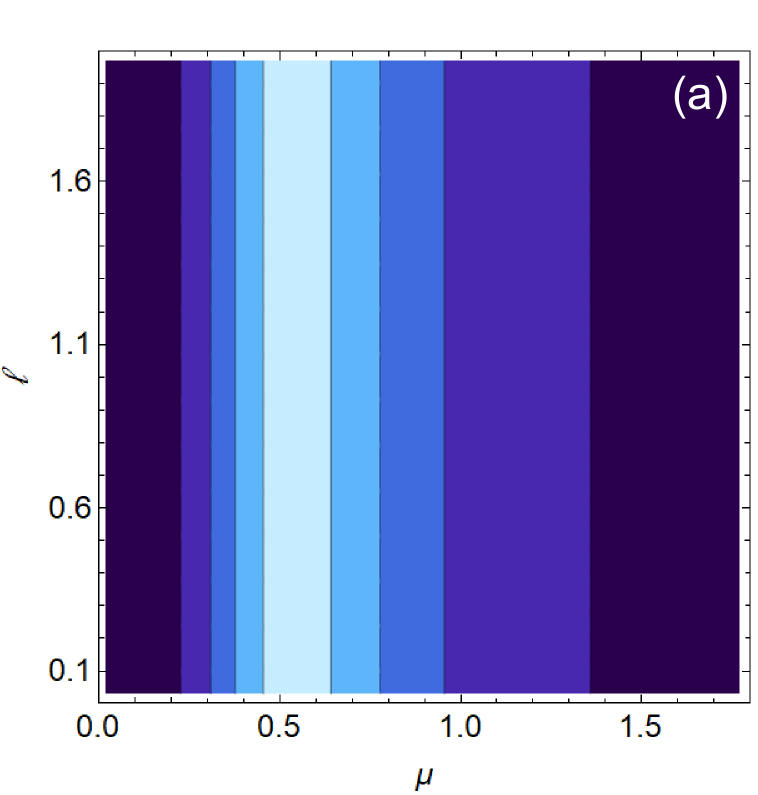}
\endminipage
\hfill
\minipage{0.48\textwidth}
\hspace{-0.9cm}
    \includegraphics[scale=0.6]{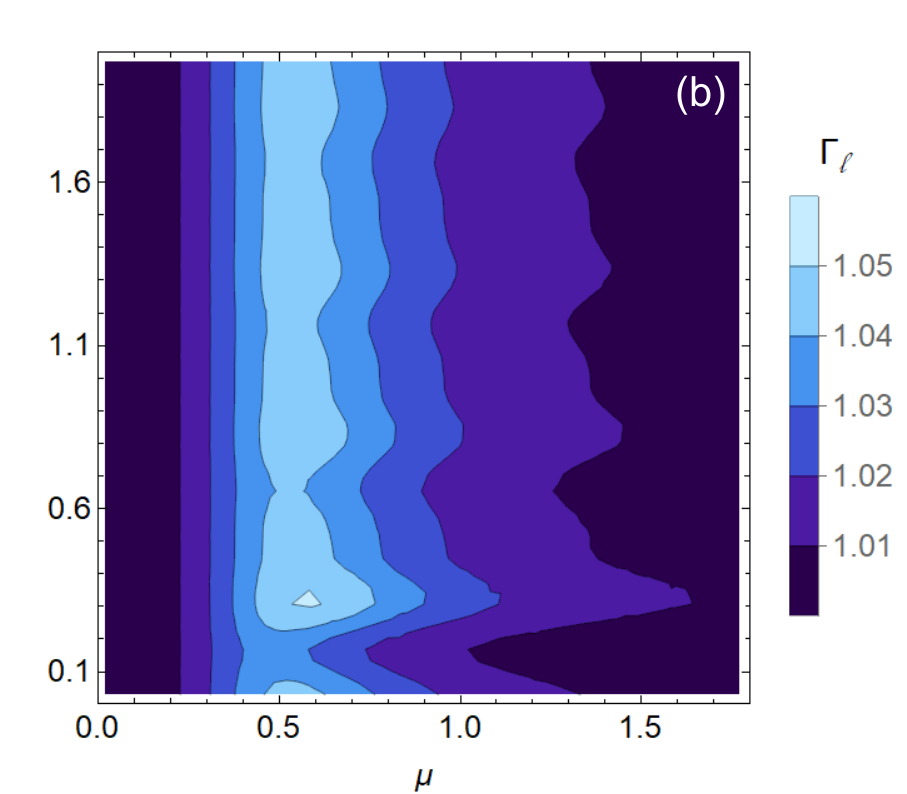}
\endminipage
\caption{Contour plot of the (a) $\ell$-independent $\Gamma$ \citep{Mendes2022} and (b) $\ell$-dependent $\Gamma_{\ell}$ from eq.~\eqref{eq:Gammal}, as a function of relative water depth with a fixed wave steepness of $\varepsilon = 1/20$.}
\label{Fig:f3}
\end{figure*}
\begin{figure*}
\centering
    \includegraphics[width=0.94\textwidth]{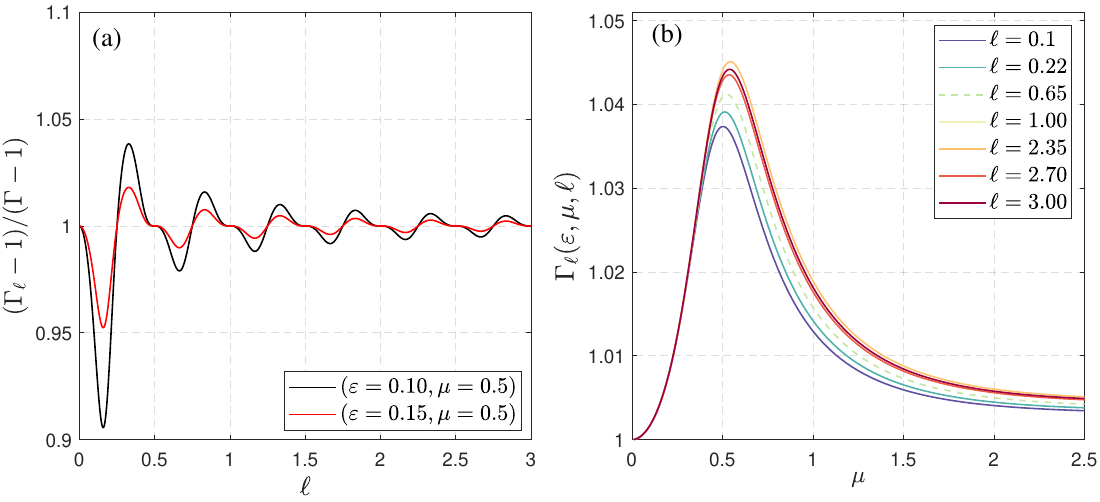}
    \caption{Effect of the non-dimensional shoaling length parameter $\ell$ on (a) the ratio of the relative non-homogeneous parameter $(\Gamma_{\ell} - 1)/(\Gamma - 1)$ and (b) the evolution of $\Gamma_{\ell}$ as a function of relative water depth $\mu =kh$, with steepness $\varepsilon=0.05$ and $\ell \in [0.1,3.0]$.}
\label{Fig:ell}
\end{figure*}
\begin{eqnarray}
\mathscr{E} =  \frac{a^{2}}{2} \left\{ \Big[  1 +  \mathcal{H}_{11\ell} \Big] + \Big[  1 + \mathcal{H}_{12\ell} \Big]  \left(  \frac{\pi \varepsilon}{4} \right)^{2} \frac{ \left( \Tilde{\chi}_{1}    +  \chi_{1} \right)}{2}     +  \frac{(1+f_{3})  }{2} \mathcal{H}_{13\ell} \frac{\pi \varepsilon}{4} \sqrt{ \Tilde{\chi}_{1} } 
\right\}  .
\label{eq:energyFIM}
\end{eqnarray}

The term $\mathcal{H}_{13\ell}(\pi \varepsilon/4) \sqrt{ \Tilde{\chi}_{1} }$ of the third-order shoaling length correction is about 1/5 of its second-order counterpart $(1+\mathcal{H}_{12\ell}) (\pi \varepsilon/4)^{2} \left( \Tilde{\chi}_{1} + \chi_{1} \right)/2$. Now, because second-order wave fields travelling from deep to intermediate water have bound values of $0.5 < f_{3} < 1.3$, the term $0.75 <(1+f_{3})/2 < 1.15$ is too small to make the third-order effect in $\ell$ comparable to the second-order one. Thus, the second-order effect due to $\mathcal{H}_{12\ell}$ has the leading order.
Accordingly, plugging eqs.~(\ref{eq:energyVAR}) and (\ref{eq:energyFIM}) into the definition of the non-homogeneous parameter $\Gamma \equiv \langle \eta^{2} \rangle / \mathscr{E}$, we arrive at:
\begin{eqnarray}
\hspace{-0.7cm}
\Gamma_{\ell} \equiv \frac{\langle \eta^{2} \rangle}{\mathscr{E}}  \approx \frac{ \left[ 1 + \frac{\sin{(4\pi \ell)} }{4\pi \ell} \right]  + \left[ 1 + \frac{\sin{(8\pi \ell)} }{8\pi \ell} \right] \left(  \frac{\pi \varepsilon \mathfrak{S} }{4} \right)^{2} \Tilde{\chi}_{1} }{   \left[ 1 + \frac{\sin{(4\pi \ell)} }{4\pi \ell} \right]  + \left[ 1 + \frac{\sin{(8\pi \ell)} }{8\pi \ell} \right] \left(  \frac{\pi \varepsilon \mathfrak{S}}{4} \right)^{2} \frac{\left( \Tilde{\chi}_{1}    +  \chi_{1} \right)}{2}  } \equiv \frac{ 1+   \frac{\pi^{2} \varepsilon^{2} \mathfrak{S}^2}{16}     \, \Tilde{\chi}_{1} \mathcal{F}(\ell) }{  1+   \frac{\pi^{2} \varepsilon^{2} \mathfrak{S}^2}{32}   \, \left( \Tilde{\chi}_{1}  + \chi_{1} \right) \mathcal{F}(\ell)   } ,
\label{eq:Gammal}
\end{eqnarray}
with $\mathcal{F}(\ell) = (1+\mathcal{H}_{12\ell})/(1+\mathcal{H}_{11\ell})$, and the vertical asymmetry $\mathfrak{S}\equiv 2H_{crest}/H$ is twice the ratio of crest height to wave height. The original definition of $\Gamma$ in \citet{Mendes2022} is now a special case of $\Gamma_{\ell}$ with $\ell=1$ by default. In \jfm{appendix}~\ref{append:Arbitrary_distrance}, we provide further justification that in the new definition of $\Gamma_{\ell}$, the integration interval involved in eq.~\eqref{eq:variance_app} and \eqref{eq:energyX0} is not necessarily from $0$ to $L$. Different integration intervals are tested, resulting in a value close to $\Gamma_{\ell}$, as long as it covers the sloping area. 

In comparison to the $\ell$-independent parameter $\Gamma$ from \citet{Mendes2022}, shown in \jfm{figure}~\ref{Fig:f3}\jfm{a}, the oscillations due to $\ell$ up to second order can be seen in \jfm{figure}~\ref{Fig:f3}\jfm{b} over a wide range of relative water depth. In shallow water with $\mu<0.5$, the oscillations due to the $\ell$ effect are negligible. Oscillations start to grow for $\mu \ge 1/2$ all the way to deep water, but $\Gamma - 1$ vanishes for large $\mu$, so eventually the impact of the oscillations due to $\ell$ effect is rather limited.

It is easily verified that $|\mathcal{F}(\ell) - 1|< 0.45$ for all $\ell$, and for $\ell > 0.5$ we have an even lower local bound $|\mathcal{F}(\ell) - 1|< 0.2$. To illustrate the minor effect of $\ell$ on the parameter $\Gamma$, \jfm{figure}~\ref{Fig:ell}\jfm{a} shows the ratio $(\Gamma_{\ell}-1)/(\Gamma-1)$ as a measure of the relative differences between computations of $\Gamma_{\ell}$ for any $\ell$ and the original $\Gamma$ (namely $\Gamma_{\ell}(\ell=1)$). Within the range from linear to second-order Stokes theory for maximum steepness of $\varepsilon \leqslant 1/7$, it shows that $\ell$ induces very minor effects on $\Gamma_{\ell}$ for $\ell >1$, and consequently on the wave statistics. For $\ell<1$, these oscillations are larger but are of little consequence for the non-homogeneous parameter $\Gamma_{\ell}-1$, since a finite spectral width and random phases in eq.~\eqref{eq:energyX0} would further reduce the amplitude of the oscillations. For instance, summing up over a phase range $[0,\pi/2]$ reduces the oscillation amplitude by 30\%. As a result, oscillations in $\Gamma_{\ell}-1$ stay below $10$\% for all $\ell$ and of 0.5\% for $\Gamma-1$, such that the effect of shoaling length $\ell$ on abnormal statistics is very small. \jfm{figure}~\ref{Fig:ell}\jfm{b} shows the evolution of $\Gamma_{\ell}$ as a function of $\mu$, with $\varepsilon=0.12$ and $\ell\in[0.1, 3]$. It is noticed that the differences among $\Gamma_{\ell}$ with various slope lengths are smaller than those due to change of relative water depth.

\subsection{Numerical simulations on arbitrary plane slopes with fixed gradient}\label{subsec:num}

\begin{figure*}
\centering
    \includegraphics[width=0.99\textwidth]{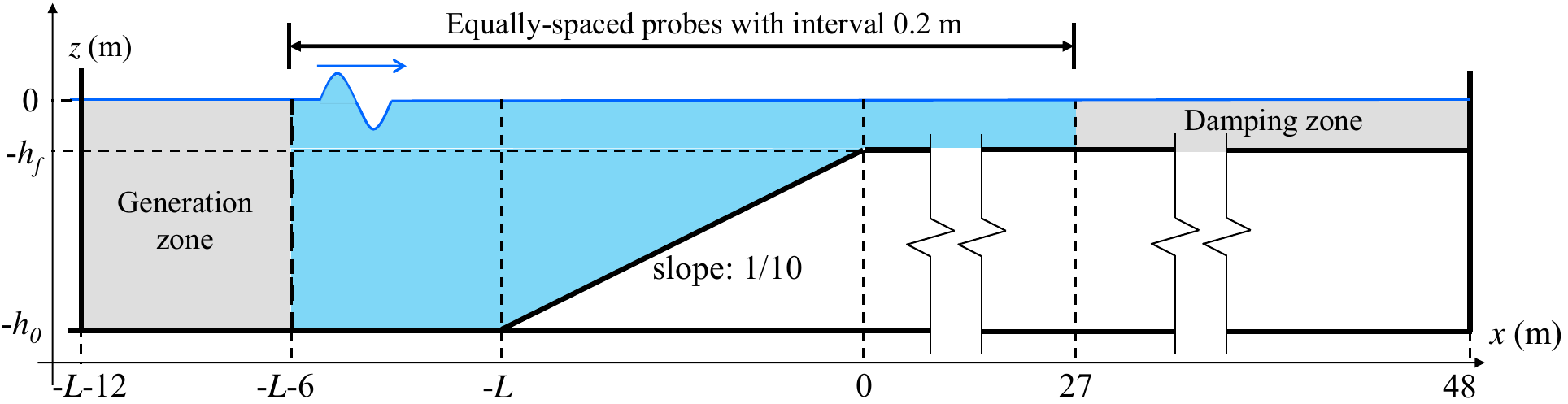}
    \caption{Sketch of the numerical wave flume with varying slope length $L$ and a constant slope gradient $(\nabla{h})_1=1/10$ (not to scale).}
\label{Fig:bathy}
\end{figure*}

To show that the negligible effect of $\ell$ on wave statistics holds for both steep and mild slopes with a constant gradient, a set of simulations is performed with a fully nonlinear potential flow solver, Whispers3D. The numerical wave flume consists of a 12~m deeper flat region, a 1/10 upslope, and a 48~m shallower flat region. The waves are generated in a 6~m generation zone and dissipated in a 21~m damping zone, see \jfm{figure}~\ref{Fig:bathy}. The shallower region has a fixed water depth $h_f=0.11$~m, while the deeper region depth $h_0$ varies depending on $L$. The bathymetry setup is very similar to the one used in \citet{Zhang2024}, but with a milder slope. The input wave fields are described by a JONSWAP spectrum with peak period $T_p=1.1$~s, and peak enhancement parameter $\gamma = 3.3$. The main wave and bottom parameters are listed in \jfm{table}~\ref{tab:cases3}, they are devised to simulate the changes in water wave parameters as a function of the shoaling length $\ell_p$ with a milder slope ($(\nabla{h})_1=1/10$) than that ($(\nabla{h})_2=1/3.8$) used in \citet{Zhang2024}.

\begin{table*}
  \begin{center}
\def~{\hphantom{0}}
  \begin{tabular}{cccccccccccc}
  \toprule
   \multirow{2}{*}{Case} & \multirow{2}{*}{$h_0$ [m]} & \multirow{2}{*}{$L$ [m]} & \multirow{2}{*}{$\ell_p$} & \multicolumn{3}{c}{Deeper region} & \multicolumn{3}{c}{Shallower region}  & \multirow{2}{*}{$\mu_0/\mu_f$} \\
   \cmidrule(r){5-7}  \cmidrule(r){8-10}
    &  &  &  & $H_{s,0}$ [m] & {$\varepsilon_0$} & {$\mu_0$} & $H_{s,f}$ [m] & {$\varepsilon_f$} & {$\mu_f$} & \\
    \midrule
   1 & 0.121 &  0.11 &  0.1 & 0.0193 & 0.0489 & {0.68} &  0.0172 & 0.0453 & {0.64} &  1.06\\
   2 & 0.131 &  0.21 &  0.2 & 0.0196 & 0.0480 & {0.71} &  0.0177 & 0.0467 & {0.64} &  1.11\\
   3 & 0.153 &  0.43 &  0.4 & 0.0201 & 0.0462 & {0.78} &  0.0185 & 0.0486 & {0.64} &  1.21\\
   4 & 0.174 &  0.64 &  0.6 & 0.0203 & 0.0442 & {0.84} &  0.0188 & 0.0494 & {0.64} &  1.31\\
   5 & 0.185 &  0.75 &  0.7 & 0.0205 & 0.0435 & {0.87} &  0.0190 & 0.0500 & {0.64} &  1.36\\
   6 & 0.217 &  1.07 &  1.0 & 0.0208 & 0.0416 & {0.97} &  0.0195 & 0.0514 & {0.64} &  1.50\\
   7 & 0.325 &  2.15 &  2.0 & 0.0210 & 0.0369 & {1.27} &  0.0197 & 0.0518 & {0.64} &  1.97\\
   8 & 0.539 &  4.29 &  4.0 & 0.0203 & 0.0318 & {1.88} &  0.0185 & 0.0489 & {0.64} &  2.92\\
   9 & 0.754 &  6.44 &  6.0 & 0.0196 & 0.0297 & {2.54} &  0.0175 & 0.0461 & {0.64} &  3.94\\
   10& 1.183 & 10.73 & 10.0 & 0.0193 & 0.0353 & {3.94} &  0.0167 & 0.0463 & {0.64} &  6.11\\
  \bottomrule
  \end{tabular}
  \caption{Summary of the key wave parameters for the simulations in the numerical wave flume with varying slope length $L$ and a constant slope gradient $(\nabla{h})_1=1/10$, the peak period $T_p=1.1$~s, and peak enhancement parameter $\gamma = 3.3$. The steepness parameter is defined as $\varepsilon \equiv (\sqrt{2}/\pi)k_pH_s$.}
  \label{tab:cases3}
  \end{center}
\end{table*}
\begin{figure*}
\centering
    \includegraphics[width=0.85\textwidth]{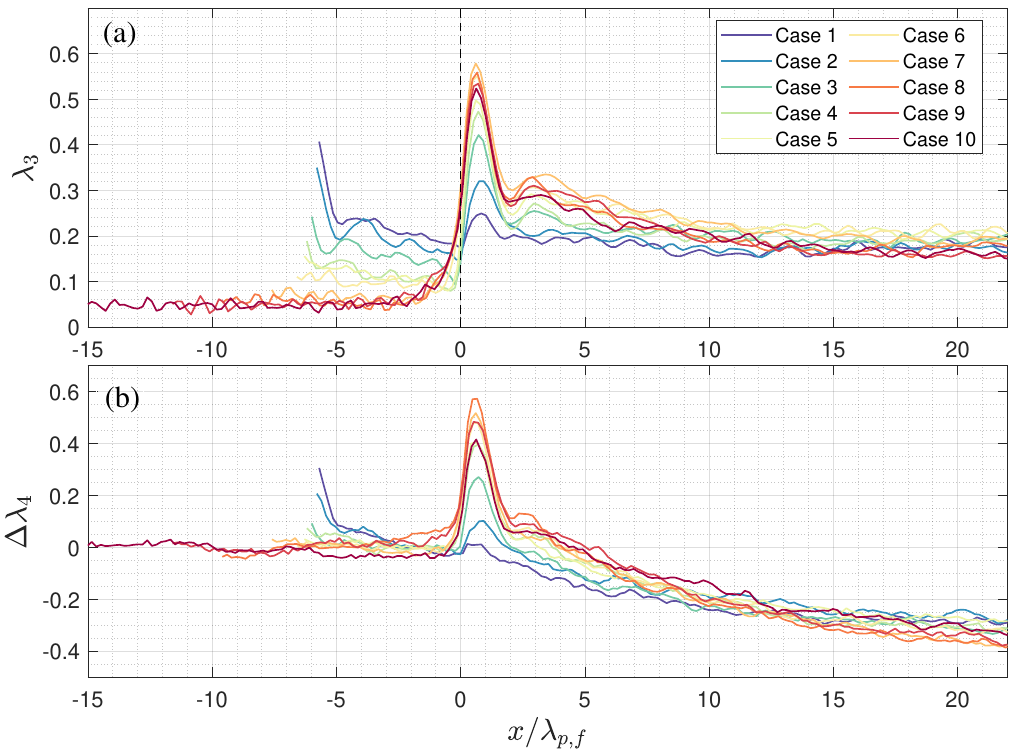}
    \caption{Comparison of the skewness $\lambda_3$ and kurtosis  enhancement $\Delta\lambda_4$ atop the shoal between simulated results for the cases of \jfm{table}~\ref{tab:cases3}. The vertical black dashed line represents the end of the shoal.}
\label{Fig:SK}
\end{figure*}
Given the similarity in wave configurations and the shared use of the Whispers3D numerical code (already introduced in \citet{Zhang2024}), the model description and numerical parameter calibration procedures are omitted here for conciseness. The duration of the simulations is 5060~s, and the FSE time series is saved every 0.2~m in the computation domain with a sampling frequency of 100~Hz. 

\jfm{Figure}~\ref{Fig:SK} shows the spatial evolution of the skewness $\lambda_3$ of FSE and the net change of kurtosis $\Delta\lambda_4$. It exhibits a behavior similar to that of a steeper slope, where the shoaling length parameter exceeds 1 ($\ell_p > 1$). In this case, the change in slope length has a negligible impact on the peak of excess kurtosis and skewness. We notice that the maximum values of the skewness and excess kurtosis are significantly lower than in the steep slope case (compare \jfm{figure}~\ref{Fig:SK} with figure 10 in \citet{Zhang2024}), which is not surprising since a milder slope means a less abrupt change of environment. Furthermore, for cases with a short slope length (cases 1 to 6), the skewness and excess kurtosis do not start from 0 but from a positive value. This is because, in the milder slope cases, a short slope length means that $h_0$ is close to $h_f$. Therefore, the linear wavemaking approach is less effective in generating nonlinear waves. Second-order irregular wave generation will be implemented in the future.

Based on the computation of $\Gamma$ parameter (originally without $\ell$ correction) and asymetry parameter $\mathfrak{S}$, the full computation for the enhancement of the kurtosis $\Delta\lambda_4$ can be computed following \citep{Mendes2023b}:
\begin{equation}
\Delta{\lambda}_{4} \approx \frac{1}{9} \left[ e^{ 8 \left( 1 - \frac{ 1 }{ \mathfrak{S}^{2}\Gamma } \right) }- 1 \right] \quad .
\label{eq:refKurt2}
\end{equation}
This model is used here with the new $\Gamma_{\ell}$ parameter to include the slope length effect explicitly. In addition, \citet{Zhang2024} has shown that combining the slope effect with a simpler general model for the kurtosis enhancement over relatively steep slopes. This model is more suitable for estimating the maximum excess kurtosis atop the shoal relative to the preshoal condition in practical applications. The simplified model reads:
\begin{equation}
\Delta{\lambda}_{4} \approx 20 \pi^{2}   \left[\left( \frac{\varepsilon_f}{\mu_f}\right)^2 -\left(\frac{\varepsilon_0}{\mu_0} \right)^{2}\right]  \sqrt{\nabla h} \quad . 
\label{eq:refKurt}
\end{equation}
Note that the steepness and relative water depth should be computed with the zero-crossing period and significant wave height. The spectral peak and thus $\mu$ remain the same, and the zero-crossing steepness can be evaluated with spectral quantities with $\varepsilon \equiv (\sqrt{2}/\pi)k_pH_s$ for irregular waves.

\begin{figure}
\centering
    \includegraphics[scale=0.7]{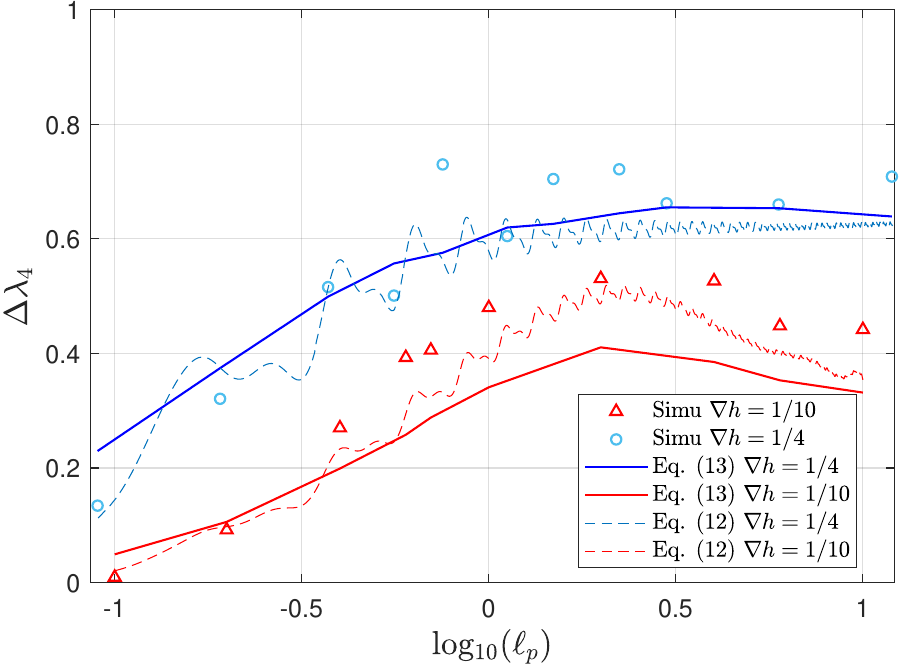}
    \caption{Comparison of the kurtosis enhancement $\Delta\lambda_4$ atop the shoal between simulated results (symbols) and theoretical estimations based on the simplified slope-dependent model of eq.~(\ref{eq:refKurt}) (solid lines) for two different slope magnitudes as well as the computation based on the full $\ell_p$ effect combining eqs.~(\ref{eq:Gammal}-\ref{eq:refKurt2}), displayed as functions of the shoaling length parameter $\ell_p$.}
\label{Fig:5}
\end{figure}
\begin{figure}
\centering
    \includegraphics[scale=0.6]{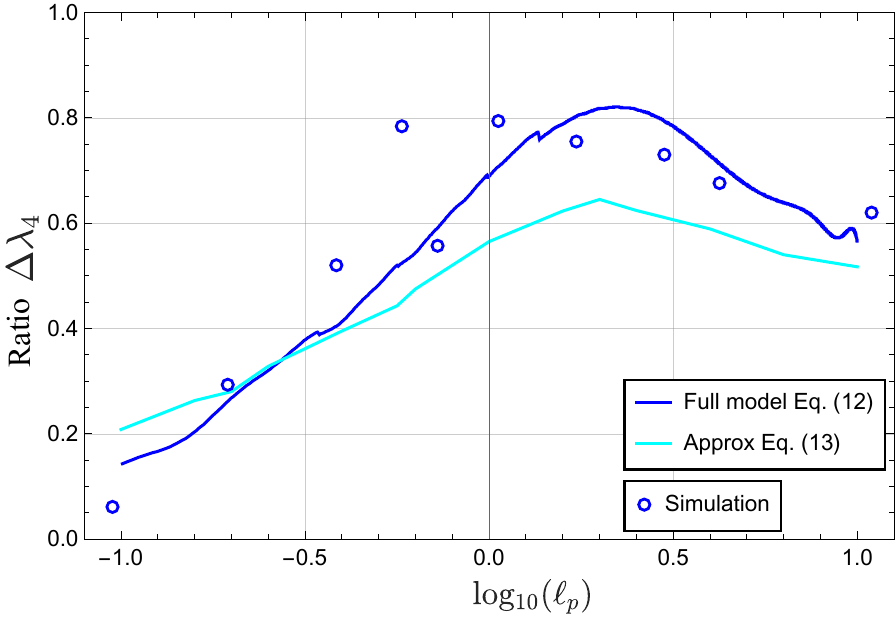}
    \caption{The ratio between the excess kurtosis of cases for bottom slopes of $1/4$ and $1/10$ as a function of shoaling length. The dashed black line depicts their ratio according to a simple function $\sqrt{(\nabla{h})_1 / (\nabla{h})_2}$ with fixed sea parameters.}
\label{Fig:6}
\end{figure}
Therefore, we can compare in \jfm{figure}~\ref{Fig:5} how the above models for the change of kurtosis describe the evolution over different shoaling lengths according to \jfm{table} \ref{tab:cases3} for two different bottom slope magnitudes, $1/3.81 \approx 1/4$ and $1/10$. 
We observe that oscillations from the model of eq.~\eqref{eq:refKurt2} seem to be relevant only for $\ell_p < 1$ for both slope magnitudes as expected from \jfm{figure}~\ref{Fig:ell}\jfm{a}. However, for $\ell_p > 1$ these oscillations seem to stop or become insignificant for both slope magnitudes. Thus, we confirm that the results of \citet{Zhang2024} hold for mild and steep slopes, i.e., the shoaling length has a limited impact on the wave statistics. However, noticeably, the model eq.~\eqref{eq:refKurt2} underpredicts the excess kurtosis between $0.4 < \ell_p < 2$ in the case of mild slopes. 
\xx{The latter can be explained if we reconsider the $\ell$ model coupled with a finite-amplitude slope-dependent theory as in \citet{Mendes2025}: as eq.~(\ref{eq:refKurt}) is a Taylor expansion of the full computation, steeper slopes will have a slightly underpredicted kurtosis due to the existing saturation of slope effect for sharp bottom gradients. Nevertheless, for such a model, we cannot simply use the superposition of terms to include $\mathcal{F}(\ell)$ attached to the super-harmonic terms, being thus beyond the scope of this work.}

\begin{figure*}
\centering
    \includegraphics[width=0.95\textwidth]{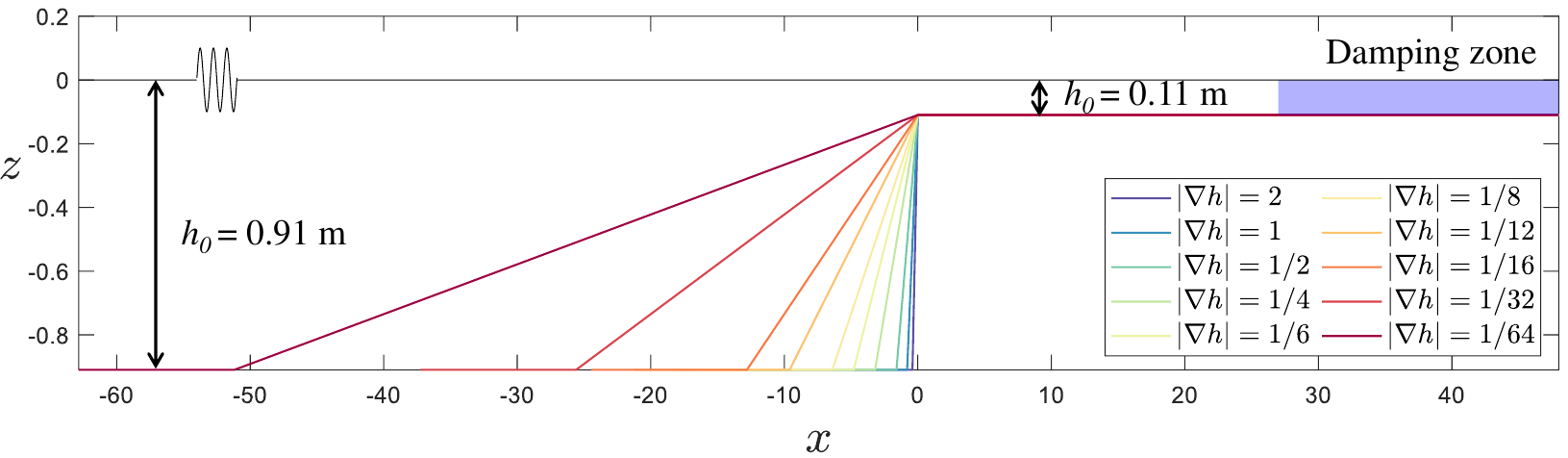}
    \caption{Sketch of the numerical wave flume with varying slope gradient $\nabla{h}$ and a constant depth transition from $h_0=0.91$~m to $h_f=0.11$~m (not to scale).}
\label{Fig:bathy2}
\end{figure*}

\begin{table*}
  \begin{center}
\def~{\hphantom{0}}
  \begin{tabular}{cccccccccccc}
  \toprule
   \multirow{2}{*}{Case} & \multirow{2}{*}{$\nabla{h}$} & \multirow{2}{*}{$L$ [m]} & \multirow{2}{*}{$\ell_p$} & \multicolumn{3}{c}{Deeper region} & \multicolumn{3}{c}{Shallower region}  & \multirow{2}{*}{$\mu_0/\mu_f$} \\
   \cmidrule(r){5-7}  \cmidrule(r){8-10}
    &  &  &  & $H_{s,0}$ [m] & {$\varepsilon_0$} & {$\mu_0$} & $H_{s,f}$ [m] & {$\varepsilon_f$} & {$\mu_f$} & \\
    \midrule
   1 & 1/0.5  &  0.4  &  0.4  & 0.0202 & 0.0241 & \multirow{10}{*}{3.04} &  0.0195 & 0.0395 & \multirow{10}{*}{0.64} &  \multirow{10}{*}{4.72}\\
   2 & 1/1    &  0.8  &  0.8  & 0.0202 & 0.0241 &  &  0.0198 & 0.0403 &   &  \\
   3 & 1/2  &  1.6  &  1.5  & 0.0197 & 0.0236 &  &  0.0193 & 0.0390 &   &  \\
   4 & 1/4  &  3.2  &  3.0  & 0.0199 & 0.0238 &  &  0.0196 & 0.0397 &   &  \\
   5 & 1/6  &  4.8  &  4.5  & 0.0204 & 0.0245 &  &  0.0201 & 0.0411 &   &  \\
   6 & 1/8  &  6.4  &  6.0  & 0.0199 & 0.0239 &  &  0.0196 & 0.0398 &   &  \\
   7 & 1/12 &  9.6  &  9.0  & 0.0198 & 0.0239 &  &  0.0195 & 0.0398 &   &  \\
   8 & 1/16 &  12.8 &  12.0 & 0.0201 & 0.0241 &  &  0.0198 & 0.0400 &   &  \\
   9 & 1/32 &  25.6 &  23.9 & 0.0199 & 0.0239 &  &  0.0196 & 0.0397 &   &  \\
   10& 1/64 &  51.2 &  47.7 & 0.0202 & 0.0239 &  &  0.0197 & 0.0400 &   &  \\
  \bottomrule
  \end{tabular}
  \caption{Summary of the key wave parameters for the simulations in the numerical wave flume with varying slope gradient $\nabla{h}\in[1/64,1/0.5]$ and a constant depth transition from $h_0=0.91$~m to $h_f=0.11$~m, The incident spectral peak period $T_p$ is fixed as 1.1~s. The table shows the mean simulated quantities in the corresponding areas (therefore, with the reflection included).}
  \label{tab:cases4}
  \end{center}
\end{table*}
In \jfm{figure}~\ref{Fig:5}, we also observe that the overall trend from the simulation is well described by eq.~(\ref{eq:refKurt}), with small deviations becoming more prominent for $0.7 \lesssim \ell_{p} \lesssim 1$ for the red solid curve (slope of $1/10$). Although these deviations are within the standard error, compatible with the total length of the simulation of $\pm 0.15$ \citep{Joanes1998}, this interval in shoaling length is also the same with strongest deviation between the simulated data and eq.~(\ref{eq:refKurt}) for the steeper slope of $1/4$. This could be understood as a weak signature of the small effect of the shoaling length $\ell_p$ on wave statistics. \jfm{Figure} \ref{Fig:ell}\jfm{a} highlights that the $\ell_p$ effect will lead to major deviations only for $\ell_p \leqslant 1$, but at the second region $0 \leqslant \ell_p \lesssim 0.6$ the kurtosis is too small, and whatever strong deviations the shoaling effect can cause will be negligible in absolute value. Thus, we can confidently assert that the slope effect ($\nabla h$) takes the lead in driving anomalous wave statistics, and the shoaling length plays a much diminished role in this process and within a very narrow range of $0.5 \lesssim \ell_p \lesssim 1$, both for steep and mild slopes.
Overall, both theoretical models \xx{(eq.~(\ref{eq:refKurt2}) and eq.~(\ref{eq:refKurt}))} correctly capture the order of magnitude of the decrease in excess kurtosis and its trend from steep to mild slopes, with the first being highly complex and offering limited advantage over the second.

A complementary view of the impact of the slope gradient $\nabla{h}$ is presented in \jfm{figure}~\ref{Fig:6}: we compute the ratio of excess kurtosis from the mild slope to the steep counterpart (averaging their $\ell_P$ since it is not fixed) and compare it with a simple dashed straight line corresponding to the ratio of slopes $\sqrt{(\nabla{h})_1 / (\nabla{h})_2} \approx 5/8$. Although the sea conditions prior to the shoal slightly vary between the two sets of simulations with different slopes, they are identical atop the shoal. This allows us to attempt to estimate of the decrease in maximum kurtosis atop the shoal as simply by $\Delta{\lambda}_{4 \, , 1}/\Delta{\lambda}_{4 \, , 2} \approx \sqrt{(\nabla{h})_1 / (\nabla{h})_2}$. Given the confidence interval on the excess kurtosis, we conclude that the square-root function of the slope is an accurate model for the decay of the bottom gradient effect as long as $\ell_p > 0.4$. For faster transitions with smaller shoaling length ($\ell_p \leqslant 0.4$), both theoretical models explain the decrease in the ratio. This can be explained by the fact that the case for which $\ell_p$ of the milder slope had smaller relative water depth $\mu$, thus nearer the region of maximum amplification of the excess kurtosis at $\mu \sim 1/2$.

\section{Effect of steep slope reflection on post-shoal wave statistics}\label{sec:validation}

\subsection{Numerical simulations on plane slopes with various gradients}\label{subsec:simu}

In the previous section, we considered relatively steep bottom slopes, under which the influence of the slope magnitude is known to have already saturated \citep{Mendes2022b}. It is known that if slopes get smaller and smaller, the net potential energy will bring down wave statistics, thereby diminishing the nonlinearity of the wave field and thus the excess in kurtosis. Here we aim to investigate how reflection affects the kurtosis change, and consequently the extreme wave statistics, over the plateau following a steep slope. 
As has been shown in \citet{Zhang2024} and discussed in the previous section, the slope length and slope gradient $\nabla{h}$ can be disentangled and investigated separately by keeping $\ell$ or $\nabla{h}$ constant. It has already been shown that the responses of skewness and kurtosis are insensitive to the slope length. Therefore, the slope gradient effect can be investigated by varying $\nabla{h}$, meanwhile, instead of keeping $\ell$ constant (as it requires a large computational domain in cases with low $\nabla{h}$), taking different $\ell$ and keeping a constant dimensional depth $h$ before the shoal. 

The new set of bathymetry is displayed in \jfm{figure}~\ref{Fig:bathy2}, and the configurations of incident wave fields are given in \jfm{table}~\ref{tab:cases4}. The water depths before and after the shoal are $h_0=0.91$~m and $h_f=0.11$~m, respectively. The slope gradient varies in $[1/64,2]$, noticing that $\nabla{h}=2$ is very steep in coastal areas. The core of this setup is to cut the slope at the deep water criterion $\mu_0\approx \pi$, and assume that the statistical properties of the incident wave fields do not modulate before entering the finite water regime. The simulations are performed with Whispers3D and with the same set of choices of numerical parameters as in section~\ref{subsec:num}, and thus are not duplicated here. Such configurations enable the investigation of how different levels of reflection rate affect the post-shoal statistics.

\begin{figure*}
\centering
    \includegraphics[width=0.9\textwidth]{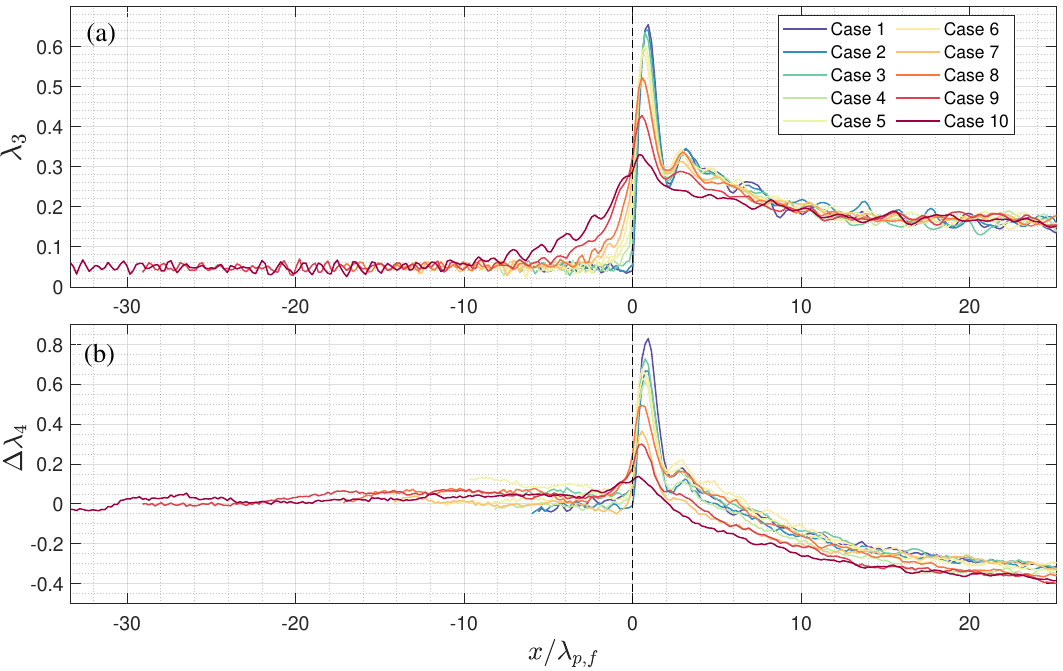}
    \caption{Comparison of the skewness $\lambda_3$ and kurtosis enhancement $\Delta\lambda_4$ atop the shoal between simulated results for the cases of \jfm{table}~\ref{tab:cases4}. The vertical black dashed line represents the end of the shoal.}
\label{Fig:SK2}
\end{figure*}

In \jfm{figure}~\ref{Fig:SK2}, the spatial evolution of skewness $\lambda_3$ and the net change of kurtosis, $\Delta\lambda_4$ in all cases listed in \jfm{table}~\ref{tab:cases4} is displayed. In \jfm{figure}~\ref{Fig:SK2}\jfm{a}, it is seen that as the slope gradient becomes steeper, the evolution of the skewness over the shoal becomes more and more abrupt. In case 10 with $\nabla{h}=1/64$, $\lambda_3$ gradually increases over the shoal and approaches the global maximum value at the end of the slope. However, in case 1 with $\nabla{h}=2$, $\lambda_3$ changes from 0.05 at the end of the slope to its global maximum value 0.64 at the beginning of the plateau in an abrupt manner. Over the plateau, the evolution trend of $\lambda_3$ is very similar, despite the difference in the maximum values. Namely, the slope gradient shows a very minor effect on the position where the maximum of $\lambda_3$ is achieved, and on the equilibrium value attained in the far field. In \jfm{figure}~\ref{Fig:SK2}\jfm{b}, the transition of kurtosis from the end of the shoal and the beginning of the plateau is rather fast in almost all cases, which is different from the observation in \jfm{figure}~\ref{Fig:SK2}\jfm{a}. It also indicates that the freak wave risk changes suddenly once waves enter the shallower region, even in cases with mild slopes. Then over the plateau, as for $\lambda_3$, $\Delta \lambda_4$ achieves maximum value at almost the same position and decreases to a similar level below 3 (less prone to the freak waves). Finally, combining with the observation in \jfm{figure}~\ref{Fig:SK} for cases with different slope lengths, we conclude that the latency (the distance between the end of the slope and the position where skewness and kurtosis achieve their maxima) is independent of the slope effect. Notably, a comparison between \jfm{figures}~\ref{Fig:SK2}\jfm{b} and \jfm{figure}~\ref{Fig:SK}\jfm{b} shows that the decrease in excess kurtosis due to slope graident is monotonic, whereas the change in kurtosis due to shoaling length is negligible for cases with $\ell_{p} > 0.5$, reinforcing the conclusion that the bottom slope has the leading role in the magnitude of the maximum excess kurtosis atop the shoal.

\begin{figure}
\centering
\minipage{0.49\textwidth}
    \includegraphics[scale=0.42]{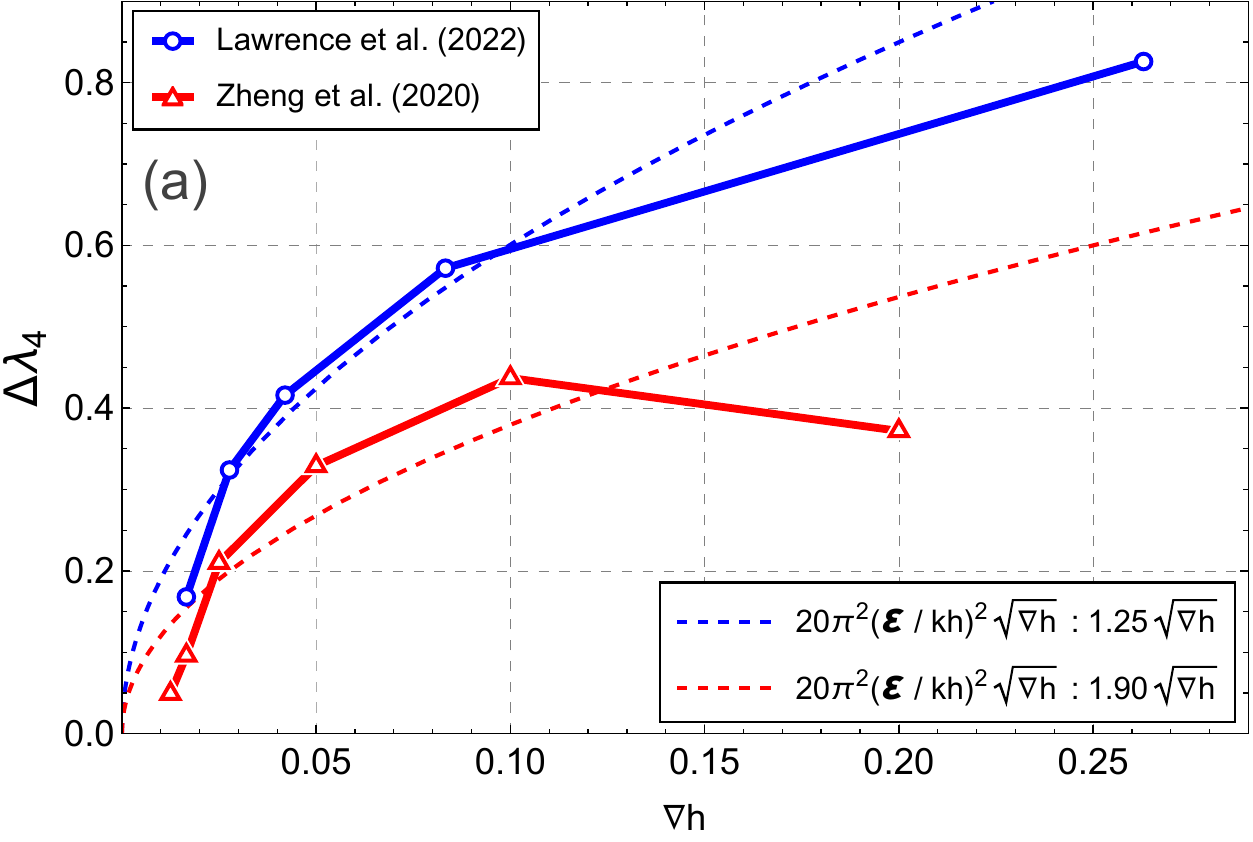}
\endminipage
\hfill
\minipage{0.49\textwidth}
    \includegraphics[scale=0.46]{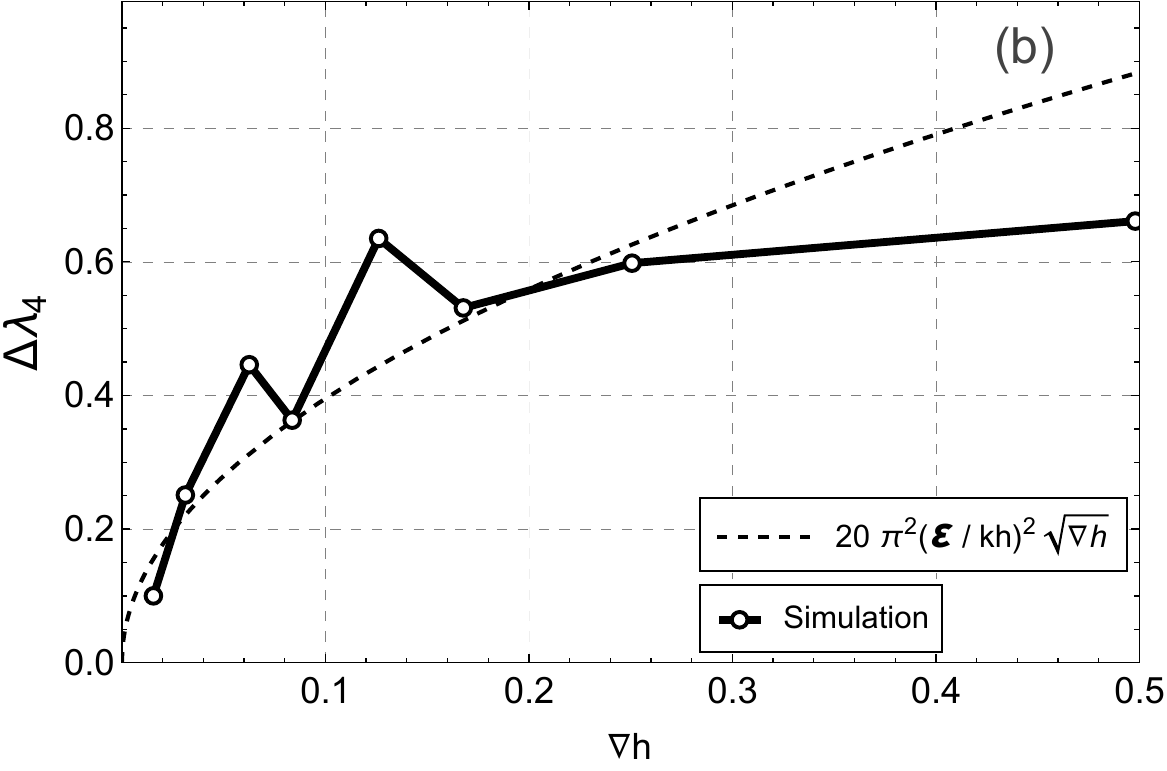}
\endminipage
\caption{\x{Trends for peak of kurtosis atop a shoal with varying bottom slope according to simulations and the model of eq.~(\ref{eq:refKurt}), panel (a) comparison of simulation results of \citet{Adcock2020} and \citet{Trulsen2022} (in solid curves) and the corresponding theoretical prediction (in dashed curves); and panel (b) comparison of simulation results of cases listed in \jfm{table}~\ref{tab:cases4} (in solid curve) and the theoretical prediction (in dashed curve);}}
\label{Fig:kurt}
\end{figure}
\begin{figure}
\centering
    \includegraphics[scale=0.6]{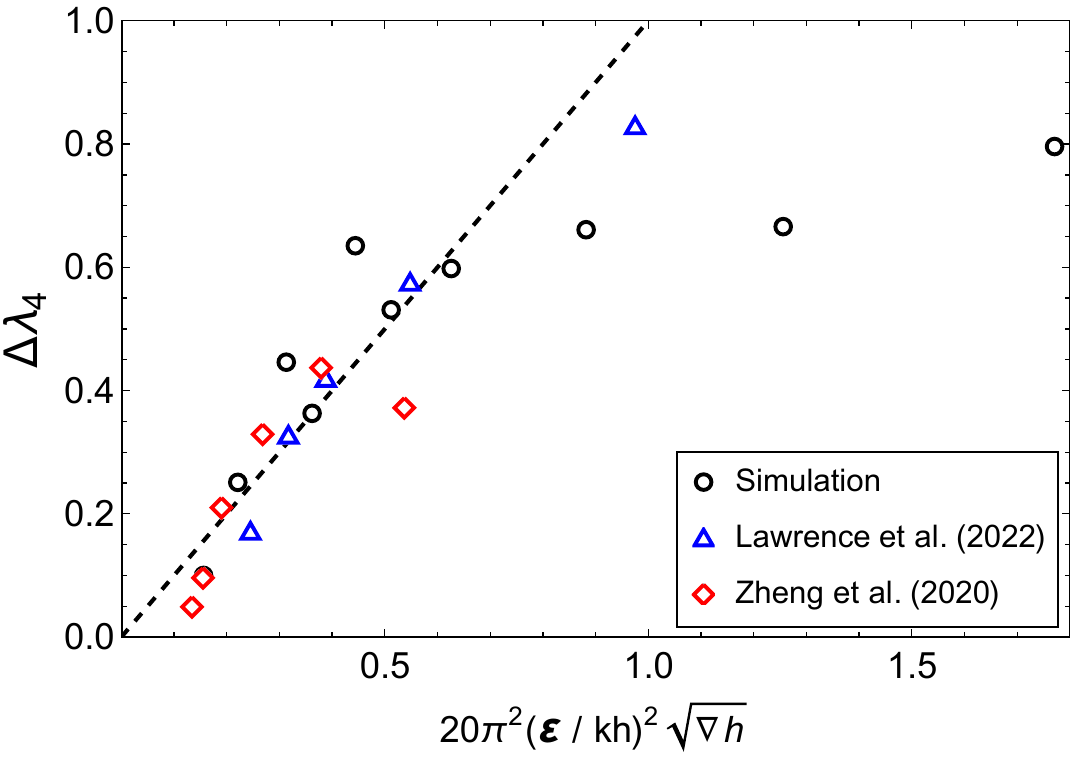}
\caption{Comparison between simulated and theoretically predicted \x{values of} excess kurtosis.}
\label{Fig:kurt2}
\end{figure}
The eq.~(\ref{eq:refKurt}) has been shown to provide a very effective model for steep slopes $\nabla h\sim 1/2$, and it will be verified here against varying $\nabla{h}$ from mild $1/64$ to very steep $1/0.5$ slopes despite being outside its validation range. 
\jfm{Figure}~\ref{Fig:kurt}a shows the comparison between the cutting-edge wave simulation results on a shoal \citep{Adcock2020} or a breakwater \citep{Trulsen2022} versus the theoretical prediction using eq.~(\ref{eq:refKurt}), and \jfm{figure}~\ref{Fig:kurt}b shows the comparison with the simulation results of the cases listed in \jfm{table}~\ref{tab:cases4}. The model of eq.~(\ref{eq:refKurt}) seems to well describe the results of the latter study, while performing reasonably well on the former. For both cases, the largest disagreement (although still small) between previous simulations and our simplified model appears to occur at slopes of $\nabla h \sim 1/4$.

The results of the simulations are provided in \jfm{figure} \ref{Fig:kurt2}. To better assess the accuracy of the model, we thus compare the simulated excess kurtosis with three different methods and sea state parameters in \jfm{figure} \ref{Fig:kurt2}. We clearly see an excellent unified description of slope-dependent kurtosis over mild and relatively steep slopes up to $\nabla h = 1/4$, upon which beyond this regime the theory overpredicts the kurtosis to some extent. Now, the standing question is why should kurtosis grow slower than $\sqrt{\nabla h}$ after slopes are too steep, noting that the saturation theory of the slope-effect \citep{Mendes2022b} is valid only up to $\nabla h \sim 1/2$. Evidently, one of the major deficiencies in dealing with bottom slope functions for water wave fields rests on the elusive mathematical treatment of reflection rates for steep slopes. In the next section, we propose a semi-analytical theory for the wave stochastics at very steep slopes that accounts for wave reflection in an attempt to forge a generalized continuous model for eq.~(\ref{eq:refKurt}) addressing any slope magnitude. 

\subsection{A semi-analytical theory for strongly reflective beaches \label{sec:steep_slopes}}

The inclusion of the reflection effect relies on the computation of the reflection coefficient. 
To better assess the impact of reflection on wave statistics, the formulae of the reflection coefficient designed for regular waves may be too simplistic. Hence, a reflection coefficient range needs to be modeled. While several formulas have been put forward for the deterministic value of reflection \citep[see][for example]{Battjes1974, Miche1951, Seelig1981}, here we intend to provide a reflection range, as a consequence of these models.

In \jfm{appendix}~\ref{append:Refl_coef}, we provide a new model of the reflection coefficient $K_R$ in a polynomial form which is \x{especially} valid for high values of surf similarity $\xi$ (for $\xi>2.5$) \x{and designed to be better suited for analytical computations of wave statistics}. \x{Since \citet{Battjes1974}, a key parameter controlling reflection rates over impermeable beaches (partially submerged slope) so-called} surf similarity, measures the slope ($\tan{\alpha} \equiv \nabla h$) compared to the steepness of the incident wave:
\begin{equation}
\xi = \frac{\nabla h}{\sqrt{\varepsilon}} \quad .
\label{eq:upxi2}
\end{equation}
Note that the definition is originally for regular waves. For irregular waves, eq.~(\ref{eq:upxi2}) measures the mean surf similarity by using the steepness $\varepsilon \equiv {H_{s,0}}/{L_{02}}$, and $L_{02}$ denotes the wavelength corresponding to the mean period $T_{02}\equiv \int_0^{\infty}E(f)\textrm{d}f/\int_0^{\infty}f^2E(f)\textrm{d}f$ of the wave field prior to the shoal \citep{Goda1975}. \x{Furthermore, it is important to note the differences and similarities between the reflection rate of a submerged slope and of a partially submerged one (beach), the former being more similar to half the section of a submerged symmetrical breakwater. Following \citet{Seelig1981}, an empirical formulation for the reflection over beaches reads:}
\begin{equation}
\x{K_R \approx \frac{a\xi^2}{\xi^2 + b} \quad ,}
\end{equation}
\x{where $a$ and $b$ are empirical coefficients that reformulate \cite{Battjes1974}'s model $0.1 \xi^2$, and supersede the former for high values of the surf similarity. Noteworthy, for rubber-mound breakwaters (rough permeable slopes) which are only partially submerged, we have $(a = 0.6, b = 6.6)$, whereas for smooth slopes (plane beaches) \citet{Seelig1981} found on average $(a = 0.6, b = 5.5)$ but with upper bound $a = 1.0$. Hence, rubber-mound breakwaters typically have a 10\% smaller reflection rate than smooth beaches, albeit this decrease may become higher up to a 40\% bound.} 

\x{On the other hand, the degree to which the breakwater is fully submerged or above the mean water level also affects reflection rates. This distance is often called the breakwater crest height, and the ratio of the former to the significant wave height has a linearly approximated effect (decreasing as the structure is submerged) on reflection rates of rubber-mound breakwaters \citep{Meer2005}. Considering empirical formulas and the test cases of \jfm{table} \ref{tab:cases4}, reflection rates would be 25\% smaller than a breakwater with null crest height, which is the limit of our bathymetry in \jfm{figure} \ref{Fig:bathy2}. However, since our simulation consists of an impermeable bottom slope, the decrease in reflection rate due to submergence is expected not to exceed 20\%.}

\x{As the deviation of the reflection rate due to a submerged slope ending in a plateau in shallow water compared to \cite{Battjes1974}'s model (in its regime of validity) is not large for the test cases of \jfm{table} \ref{tab:Ksigma}, o}ur \x{conservative} model for the reflection coefficient is formulated as follows:

\begin{figure*}
\centering
   \includegraphics[scale=0.5]{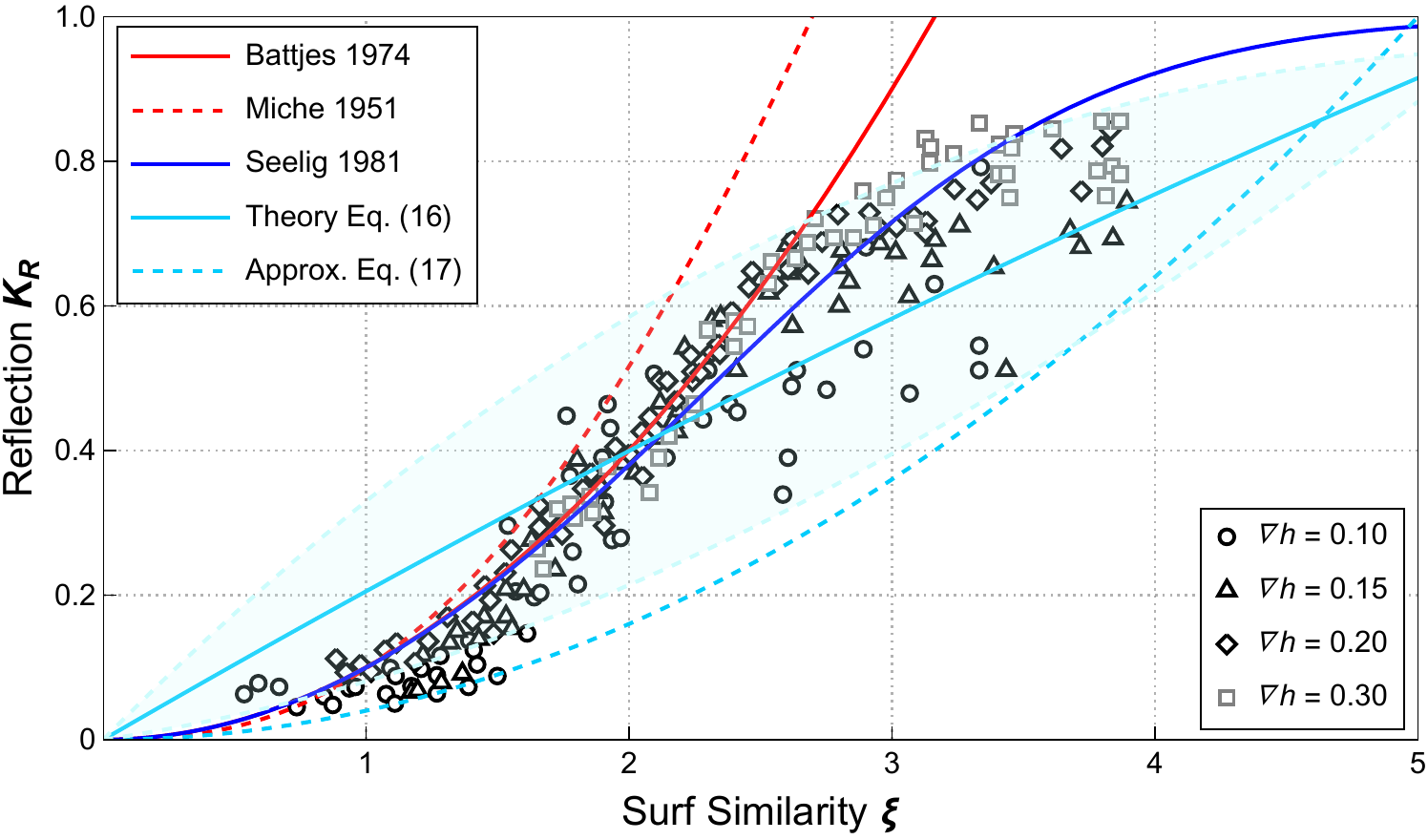}
   \caption{Comparison between several analytical models for the reflection coefficient versus experimental data collected by \citet{moraes1970experiments}. The shaded blue area corresponds to the error described by eq.~(\ref{eq:Kstd}) applied to the new model of eq.~(\ref{eq:OurMODEL_main}). The dashed cyan curve depicts the approximation of eq.~(\ref{eq:OurMODEL2_main}).}
   \label{fig:Miche4}
\end{figure*}
\begin{equation}
K_R (\xi)  \approx  \left( \frac{6 }{\xi \sqrt{\varepsilon}} - 1 \right) \frac{\xi^2}{180}  \;\; , \quad \textrm{for }\xi>2.5
\label{eq:OurMODEL_main}
\end{equation}
The reflection rate may vary for high values of surf similarity, and we approximate the lower bound as:
\begin{equation}
K_{R}^{(-)}  \approx   \left( \frac{3 }{2 \sqrt{\varepsilon}} - 1 \right) \frac{\xi^2}{180} \quad .
\label{eq:OurMODEL2_main}
\end{equation}
Remarkably, although eq.~(\ref{eq:OurMODEL_main}) is expected to be valid only for $\xi > 2.5$, it actually performs reasonably well across the entire range when combined with the upper and lower bounds $K_{R}^{(\pm)}$, as is shown in \jfm{figure}~\ref{fig:Miche4}. Whenever it overpredicts reflection rates for low surf similarity, the rates are in absolute values small. This somewhat small deviation bears little impact on wave statistics as it belongs to the regime of eq.~(\ref{eq:refKurt}). Therefore, we have uncovered a new polynomial model for the reflection rate that performs nearly as well as the hyperbolic function of \citet{Seelig1981} but also effectively acts as a lower bound for the reflection rate at high surf similarity, while an upper bound for low surf similarity. \x{The main advantage of this new approximation is that it allows us to bound maximum and minimum reflection rates surrounding the mean reflection, thus providing a reflection range for wave statistics, otherwise unfeasible with \cite{Seelig1981}'s model and improvements thereof (see the strong scatter of \jfm{figure}~\ref{fig:Miche4}).}

\subsection{Implications to Wave Statistics}\label{subsec:reflecWS}

The excess kurtosis $\hat{\lambda}_{4}$ can be computed as the product between steep slope and saturated wave statistics part $\hat{\lambda}_{4, \, b}$ and a slope-dependent part $e^{ -48 \check{\mathscr{E}}_{p2} }$, as has been demonstrated in \citet{Zhang2024},
\begin{equation}
\hat{\lambda}_{4} =  \hat{\lambda}_{4, \, b} \,   e^{ -48 \check{\mathscr{E}}_{p2} } \,\, .
\end{equation}
In the slope-dependent part, $\check{\mathscr{E}}_{p2}$ denotes the non-dimensional net change of potential energy due to set-down over the shoal: 
\begin{equation}
 \check{\mathscr{E}}_{p2} = \frac{8}{ \mathfrak{S}^{2} h_{0}^{2} } \cdot \frac{1}{\lambda}\int_{0}^{\lambda}   \left[ \langle \eta \rangle^{2}  + 2 \langle \eta \rangle h(x) \right] \textrm{d}x  \quad .
\end{equation}

\citet{BMei2005} portrays the reflection problem as measuring the surface elevation as (using $\eta_0$, $\eta_{\textrm{refl}}$, and $\eta$ to denote the incident, reflected, and the total wave field, respectively),
\begin{equation}
\eta \equiv \eta_0 + \eta_{\textrm{refl}} = A_{-} e^{-ik_1x} + A_{+} e^{ik_1x} \,\, \therefore \,\, K_R = \frac{A_{+}}{A_{-}} \quad ,
\end{equation}
where $A _-$ and $A_+$ denote the complex envelopes of the incident and reflected wave field. For wave statistical purposes, according to our framework, energetics is the leading integral property that delineates statistical distributions. As such, even if there are phase shifts due to the scattering, for energetic purposes we can use the energetic correction of the net change of potential energy $\check{\mathscr{E}}_{p2 \, , \, R}$ without boundary terms \citep{Mendes2022b}:
\begin{eqnarray}
\nonumber
\check{\mathscr{E}}_{p2 \, , \, R} &=& \frac{8}{ \mathfrak{S}^{2} h_{0}^{2} } \cdot \frac{1}{\lambda}\int_{0}^{\lambda}   \left[ \langle \eta \rangle^{2}  + 2 \langle \eta \rangle h(x) \right] dx  \quad ,
\\
&\approx& \frac{16\langle \eta_0 \rangle}{ \mathfrak{S}^{2} h_{0} } (1+K_R) \int_{0}^{\lambda} \left[ 1 + \frac{ x \nabla h }{ h_{0} }  \right] \frac{dx}{\lambda}  \quad ,
\end{eqnarray}
Since \x{the reflection coefficient can be rewritten as} $K_R \approx \tilde{\mathfrak{B}} (\nabla h)^2/{\varepsilon}$ (where $\mathfrak{B}$ denotes a reflection coefficient correction for steep slopes whose definition is provided in eq.~\eqref{eq:NEWreflec} in \jfm{appendix}~\ref{append:Refl_coef}), generalizing \cite{Battjes1974}'s model, the remaining coefficients and integration are already known and described by section III of \citet{Mendes2022b}, which brings us to:
\begin{eqnarray}
\check{\mathscr{E}}_{p2 \, , \, R} \approx  - \frac{5\varepsilon^{2}}{\mu^2} \nabla h \left[ 1 -  \nabla h + \frac{\tilde{\mathfrak{B}}}{\varepsilon} (\nabla h)^2  \right]  \,\, .
\end{eqnarray}

\x{Indeed}, we can approximate the term $\tilde{\mathfrak{B}}/\varepsilon$ with a single number following the lower bound for the reflection coefficient in eq.~(\ref{eq:OurMODEL2}). Since the model of eq.~(\ref{eq:refKurt}) already verifiably applies to slopes smaller than $\nabla h \leqslant 1/2$, strongly reflective beaches here means very steep slopes $\nabla h \gg 1/2$. For wave steepness of the order of $\varepsilon = 0.04$ and the targeted slopes discussed in \jfm{table} \ref{tab:cases4} ($1/64 \leqslant \nabla h \leqslant 2$), it implies a surf similarity bound of the order of $ \xi_{\infty} \approx 8$, and one has $ K_{R}^{(-)} \sim \xi^2/75$. Furthermore,  $\xi^2 \approx 20 (\nabla h)^2$ which lets us conclude that $K_{R}^{(-)}  \sim 4(\nabla h)^2/15$. Bearing in mind the empirically-driven confidence interval in eq.~(\ref{eq:Kstd}) has to be adapted to an approximately twice larger surf similarity and that $K_{R}^{(-)}$ is the equivalent to the reflection rate diminished by two standard deviations, 

\begin{eqnarray}
 K_{R}  \left( 1 - \frac{2\sigma}{\langle K_{R} \rangle}   \right) \approx  K_{R}^{(-)}   \leqslant K_{R}  \leqslant K_{R}^{(+)} \approx K_{R}  \left( 1 + \frac{2\sigma}{\langle K_{R} \rangle}   \right) \quad ,
\end{eqnarray}
the simplest closed-form polynomial expression for the range in which the reflection coefficient will lie with 95\% confidence as a function of slope is approximately:
\begin{eqnarray}
\frac{4}{15} (\nabla h)^2  \leqslant K_{R} \approx \frac{3}{10} (\nabla h)^2 \left[ \frac{ 8+\sqrt{20}  }{ 4 +   \sqrt{20}  \nabla h }   \right] \leqslant \frac{1}{16\pi} (\nabla h)^2 \left[ \frac{ 20+7\sqrt{20}  }{ 4 +   \sqrt{20}  \nabla h }   \right]^2 \quad ,
\label{eq:KRFin}
\end{eqnarray}
\x{where by comparison the term $\tilde{\mathfrak{B}}/\varepsilon$ represents all coefficients and functions attached to $(\nabla h)^2$ in eq.~(\ref{eq:KRFin}).}
By neglecting the reflection coefficient, the approximation of eq.(\ref{eq:refKurt}) was valid up to $\nabla h = 1/2$.
Then, if we add the reflection term to the method for the modeling of $\check{\mathscr{E}}_{p2, \;R}$ in section 2.2 of \citet{Zhang2024}, one finally gets:
\begin{equation}
\check{\mathscr{E}}_{p2 \, , \, R} \approx \frac{\varepsilon^{2}}{4\mu^2} \left[ 7 - 20 \sqrt{ \nabla h \left[1 -\nabla h + \frac{\tilde{\mathfrak{B}}}{\varepsilon} (\nabla h)^2 \right] }   \right]    \quad ,
\end{equation}
thus generalizing the kurtosis of eq.~(\ref{eq:refKurt}) with $\tilde{\mathfrak{B}}/\varepsilon$ being calculated from eq.~(\ref{eq:KRFin}):
\begin{equation}
\hat{\lambda}_{4}^{\pm} \approx 20 \pi^{2}   \left( \frac{\varepsilon}{\mu} \right)^{2}  \sqrt{\nabla h \left[1 -\nabla h + \frac{\tilde{\mathfrak{B}}}{\varepsilon} (\nabla h)^2 \right]} \approx 20 \pi^{2}   \left( \frac{\varepsilon}{\mu} \right)^{2}  \sqrt{\nabla h  -(\nabla h)^2 +  \frac{3}{10}  \left( \frac{ 8+\sqrt{20}  }{ 4 +   \sqrt{20}  \nabla h }   \right) (\nabla h)^3 }
\quad . 
\label{eq:FIMkurt}
\end{equation}
Given that the semi-analytical relationship $\hat{\lambda}_4 = (4\hat{\lambda}_3/3)^2 $ between kurtosis and skewness \citep{Mori1998} remains valid for high reflection regimes, we find the skewness to be:
\begin{equation}
\hat{\lambda}_{3}^{\pm} \approx \frac{10\pi}{3}   \left( \frac{\varepsilon}{\mu} \right)  \sqrt[4]{\nabla h \left[1 -\nabla h + \frac{\tilde{\mathfrak{B}}}{\varepsilon} (\nabla h)^2 \right]} \approx \frac{10\pi}{3}   \left( \frac{\varepsilon}{\mu} \right)  \sqrt[4]{\nabla h  -(\nabla h)^2 +  \frac{3}{10}  \left( \frac{ 8+\sqrt{20}  }{ 4 +   \sqrt{20}  \nabla h }   \right) (\nabla h)^3 } \quad .
\label{eq:FIMskew}
\end{equation}

\begin{figure*}
\centering
   \includegraphics[scale=0.7]{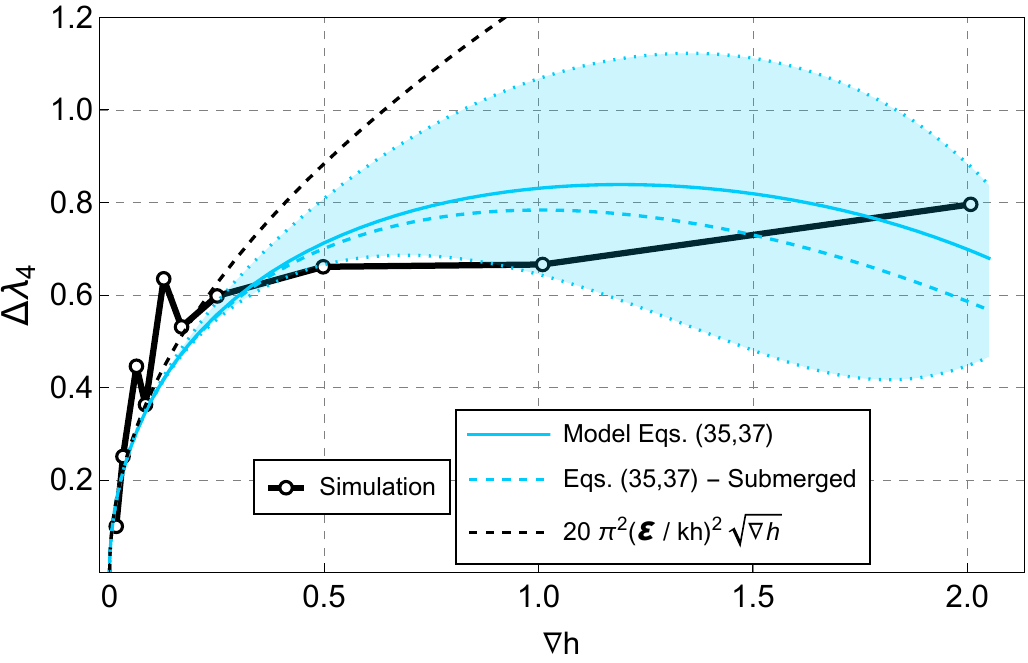}
\caption{Excess kurtosis as a function of the bottom slope magnitude. The boundary of the red shared area corresponds to the kurtosis, taking into account the reflection rate lower bound $K_{R}^{(-)}$, the solid red curve represents the adjusted mean reflection rate equivalent. The dashed curve represents a correction of -20\% on the reflection rate due to the fact that the plateau atop the shoal is submerged.}
\label{fig:kurtFIN}
\end{figure*}
\begin{figure*}
\centering
   \includegraphics[scale=0.58]{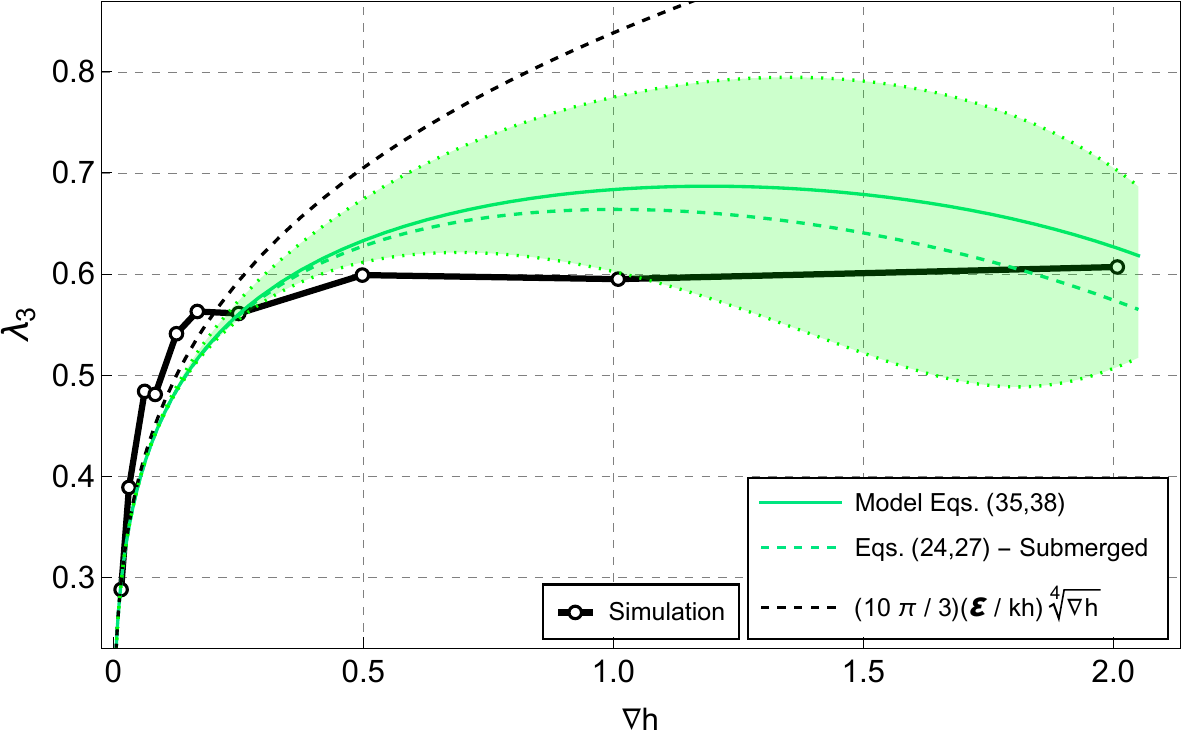} 
\caption{Skewness plotted as a function of the bottom slope magnitude with the method for the kurtosis. \x{The dashed curve represents a correction of -20\% on the reflection rate due to the fact that the plateau atop the shoal is submerged.}}
\label{fig:kurtFIN2}
\end{figure*}

In \jfm{figure} \ref{fig:kurtFIN}, we show the maximum value of $\Delta \lambda_4$ as a function of the bottom gradient $\nabla{h}$, the theoretical prediction without reflection in eq.~\eqref{eq:refKurt} in black dashed line and the new formula developed in this work for steep reflective slopes, namely eq.~(\ref{eq:FIMkurt}), in blue dashed/solid lines are superimposed. The blue area indicates the effect of reflection coefficient $K_{R}$, which may vary between $[K_{R}^{(-)}, K_{R}^{(+)}]$ and lead to a 95\% confidence interval of kurtosis increasing and later stabilizing with $K_R$. It shows that, for low slope gradients, the uncertainty in the reflection coefficient is not large, and the excess kurtosis falls within the uncertainty range of the theoretical prediction even in cases with very steep slopes. Considering that the term $\nabla h (1-\nabla h)$ is very concave, from this point of view, the very reflective beach adds another term to the net potential energy such that it stabilizes the excess kurtosis past slopes of $\nabla h \gtrsim 1/2$. Likewise, the predicted curve for the skewness according to eq.~(\ref{eq:FIMskew}) is also in good agreement with the simulations, as depicted in \jfm{figure} \ref{fig:kurtFIN2}.

\section{Conclusion \label{sec:conclusion}}

In this work, by conducting theoretical analysis as well as numerical simulations with a fully nonlinear potential flow solver, Whispers3D, we have fully probed the relationship between shoaling length, bottom slope, and reflection rates in affecting the stochastic behavior of irregular waves traveling past a plane slope. For the effect of shoaling length, we developed the non-homogeneous parameter $\Gamma_{\ell}$ with the correction due to an arbitrary slope length, which stands as a generalized form of that defined in \citet{Mendes2022}. We show that the effects of the shoaling length parameter $\ell$ and integration interval $[-D,D]$ involved in the definition only result in oscillations in $\Gamma_{\ell}$ parameter with very limited amplitude. Our simulation results confirm that the effect of shoaling length is negligible for both mild and steep slopes alike, extending previous knowledge limited to steep slopes in \citet{Zhang2024}. \xx{However, the latter effect leaves a weak signature when the wavelength is about the same as the shoaling horizontal distance. Furthermore, this weak signature is found to appear only because of finite-amplitude corrections to the oscillations caused by the shoaling length for $\ell<1$. When the effect of shoaling length is simulated with different but fixed bottom slopes, the ratio between the excess kurtosis of larger to smaller slope is stable for the most part, except for distances much larger than the wavelength. Hence, we attribute the decrease of the excess kurtosis to a smaller slope gradient, which drives the distinct exploration of the slope gradient.} 

Given that the slope length is insignificant to the net change of excess kurtosis, it is then logical to vary the bottom gradient and slope length at the same time and attribute the results solely to the bottom gradient. Therefore, in this work, a new set of bottom configurations is chosen with fixed water depths before and after the slope, and varying the slope gradient. Again, in this setup, $\nabla{h}$ and $\ell$ are both changed, but only $\nabla{h}$ plays a dominant role. The simulated results reveal that the response of wave characteristics such as excess kurtosis and skewness to an ever-increasing bottom slope is monotonic, until the plane slope is sufficiently steep $\nabla{h}\sim1/2$ and then the wave statistics reaches a regime of saturation. It is anticipated that the saturation of the increase of skewness and kurtosis is due to a high reflection coefficient. To quantify the role played by reflection, a new reflection coefficient formula in a polynomial form is put forward, which performs with similar accuracy as the model of \citet{Seelig1981} and with minimum and maximum bounds in addition. Then we include the new reflection model in the definition of non-homogeneous parameter and quantify its confidence range due to the variation of reflection rate. Comparing the simulation results with the theory for steep and reflective slopes, it is noticed that the theoretical excess kurtosis stabilizes for steep slopes with high reflection rate, and that the simulated kurtosis remains in the confidence interval of our new theory. We therefore conclude that the high reflection rate is the main reason for anomalous wave statistics becoming stable. Additionally, in the simulations, the excess kurtosis atop the shoal becomes essentially flat for plane beaches steeper than $\nabla{h}\sim1/2$, this is also well captured by our theory.

\section*{Declaration of competing interest}
The authors declare that they have no known competing financial interests or personal relationships that could have appeared to influence the work reported in this paper.

\section*{Acknowledgements}
J. Z. acknowledges the financial support of the National Natural Science Foundation of China (Grant No. 52101301), and the China Postdoctoral Science Foundation (Grants No. 2023T160078, 2021M690523).

\appendix
\renewcommand{\theequation}{\Alph{section}.\arabic{equation}}
\setcounter{equation}{0} 
\section{Auxiliary coefficients\label{append:auxilliary_coef}}

\begin{numcases}{}
\mathcal{I}_{1} \equiv \int_{0}^{L}  \cos^{2}{\phi} \, \frac{\textrm{d}x}{L} = \frac{1}{2} \left[ 1 + \frac{ \sin{(4\pi \ell)} }{4\pi \ell} \right] \, , \notag\\ \mathcal{I}_{2} \equiv \int_{0}^{L}  \cos^{2}{(2\phi)} \, \frac{\textrm{d}x}{L} = \frac{1}{2} \left[ 1 + \frac{ \sin{(8\pi \ell)} }{8\pi \ell} \right],
\label{eq:setint}
\\
\mathcal{I}_{12} \equiv \int_{0}^{L}  \cos{\phi} \, \cos{(2\phi)} \, \frac{\textrm{d}x}{L} = \frac{ 3\sin{(2\pi \ell)} +  \sin{(6\pi \ell)}}{12\pi \ell},  \notag\\
\mathcal{I}_{34} \equiv \int_{0}^{L}  \sin{\phi} \, \sin{(2\phi)} \, \frac{\textrm{d}x}{L} = \frac{  \sin^{3}{(2\pi \ell)}}{3\pi \ell}. \notag
\end{numcases}

The zero-order integrals containing $\cos{\phi} \cos{(2\phi)}$, such as $(\mathcal{I}_{12}, \mathcal{I}_{34})$, are not of leading order and are related to the mean water level correction. As in the case of assuming $\ell = 1$ in \citet{Mendes2022}, the squared trigonometric functions leading to $(\mathcal{I}_{1}, \mathcal{I}_{2})$ take the leading role in driving anomalous statistics because they still return non-zero values when the shoaling length integrals $(\mathcal{I}_{12}, \mathcal{I}_{34})$ vanish at half-integers $\ell = (2n+1)/2$, with $n \in \mathbb{N}$.

\begin{numcases}{}
\mathcal{H}_{11\ell} = \frac{\sin{(4\pi \ell)} }{4\pi \ell}, \notag\\
\mathcal{H}_{12\ell} =  \frac{\sin{(8\pi \ell)} }{8\pi \ell}, \\
\mathcal{H}_{13\ell} = \frac{3\sin{(2\pi \ell)} +  \sin{(6\pi \ell)}}{3\pi \ell} \notag.
\label{eq:H13l}
\end{numcases}

\begin{numcases}{}
\label{eq:setint2}
\mathcal{J}_{1} \equiv \int_{-h}^{0}  \cosh^{2}{\varphi} \, \textrm{d}z = \frac{h}{2} + \frac{ \sinh{(2  \mu )} }{4k}, \,\, \notag\\ \mathcal{J}_{2} \equiv \int_{-h}^{0}  \cosh^{2}{(2\varphi)} \, \textrm{d}z = \frac{h}{2} + \frac{ \sinh{(4  \mu )} }{8k}, \notag\\
\mathcal{J}_{12} \equiv \int_{-h}^{0}  \cosh{\varphi} \, \cosh{(2\varphi)} \, \textrm{d}z = \frac{ 3\sinh{ \mu } +  \sinh{(3 \mu )}}{6k},  \\
\mathcal{J}_{3} \equiv  \int_{-h}^{0}  \sinh^{2}{\varphi} \, \textrm{d}z = - \frac{h}{2} + \frac{ \sinh{(2  \mu )} }{4k}, \, \notag\\ \mathcal{J}_{4} \equiv \int_{-h}^{0}  \sinh^{2}{(2\varphi)} \, \textrm{d}z = - \frac{h}{2} + \frac{ \sinh{(4  \mu )} }{8k}, \notag\\
\mathcal{J}_{34} \equiv \int_{-h}^{0}  \sinh{\varphi} \, \sinh{(2\varphi)} \, \textrm{d}z = \frac{  2\sinh^{3}{ \mu }}{3k}.  \notag
\end{numcases}

\setcounter{equation}{0}
\section{Computation of the non-homogeneous parameter for arbitrary distances}\label{append:Arbitrary_distrance}

To explicitly evaluate the relevance of the shoaling length parameter $\ell$ on the change of the non-homogeneous parameter $\Gamma$ and the associated change in $\lambda_4$ over the shoal, we recall that the pre-shoal value of $\Gamma$ is independent on the shoal. Moreover, in the absence of wave breaking, we know that the wave steepness $\varepsilon$ and the relative water depth $\mu$ do not change significantly after the end of the shoal. Therefore, comparing changes in $\Gamma$ as a function of $L$ amounts to comparing the corresponding values of $\Gamma$ at the end of the shoal ($x = 0$). Earlier in \jfm{section} \ref{sec:ell0} detailed calculations are presented for the interval $[-L,0]$. We want to show that integration over a more general interval $[-D,D]$ for the computation of $\Gamma$ will converge to the results of $[-L,0]$. We calculate the latter by performing averages over an arbitrary spatial range $[-D, D]$ that includes the shoal interval $[-L,0]$, and with $D \gg \lambda$:
\begin{eqnarray}
\hspace{-0.3cm}
 \Gamma_d ( x = 0)  = \frac{ \frac{1}{2D} \int_{-D}^{+D}  \eta^{2} \textrm{d}x }{ \frac{1}{4D} \int_{-D}^{+D}  \eta^{2} \textrm{d}x + \frac{1}{4g D} \int_{-D}^{+D} \int_{-h(x)}^{0} \left( u^2 + w^2 \right) \, \textrm{d}z \, \textrm{d}x } \, .
 \label{eq:energyGEN}
 \end{eqnarray}
The integrals in eq.~(\ref{eq:energyGEN}) can be split into smaller intervals of integration to let $L$ appear explicitly:
 \begin{eqnarray}
 \hspace{-0.6cm}
\Gamma_d ( x = 0) = \frac{ \frac{2g}{2D}\left[ \int_{-D}^{-L}  + \int_{-L}^{0} + \int_{0}^{D}  \right]  \eta^{2}  \textrm{d}x }{ \frac{g}{4D} \left[ \int_{-D}^{-L}  + \int_{-L}^{0} + \int_{0}^{D}  \right]  \eta^{2}  \textrm{d}x + \frac{1}{4D} \left[ \int_{-D}^{-L}  + \int_{-L}^{0} + \int_{0}^{D}  \right] \int_{-h}^{0} \left( u^2 + w^2 \right) \, \textrm{d}z \, \textrm{d}x } \, ,  
\label{eq:energyGEN2}
\end{eqnarray}
As delineated in eq.~(\ref{eq:energyGEN2}), the interval can be split into three regions (I, II, III), covering the subsets $[-D,-L], [-L,0] $ and $[0,D]$. The main integrals of \jfm{section} \ref{sec:ell0} can be computed for the other regions as well:
\begin{eqnarray}
 \int_{-D}^{-L}  \cos^{2}{\phi} \, \frac{\textrm{d}x}{(D-L)}  = \frac{1}{2} \left[ 1 + \frac{ \sin{(4\pi d)} - \sin{(4\pi \ell)} }{4\pi (d -\ell)} \right] \, ,  \,
\int_{0}^{D}  \cos^{2}{(2\phi)} \, \frac{\textrm{d}x}{D}  =  \frac{1}{2} \left[ 1 + \frac{ \sin{(8\pi d)} }{8\pi d} \right] \, , \,  d 
 \equiv \frac{D}{\lambda}\,\, .  
\label{eq:setint00}
\end{eqnarray}
As seen in eq.~(\ref{eq:Gammal}), the calculation leading to the spatial corrections on $\Gamma$ is ultimately encoded in the terms $\mathcal{H}_{11},\mathcal{H}_{12}$. Thus, these coefficients can be summarized for all three regions:
\begin{figure*}
\centering
    \includegraphics[scale=0.57]{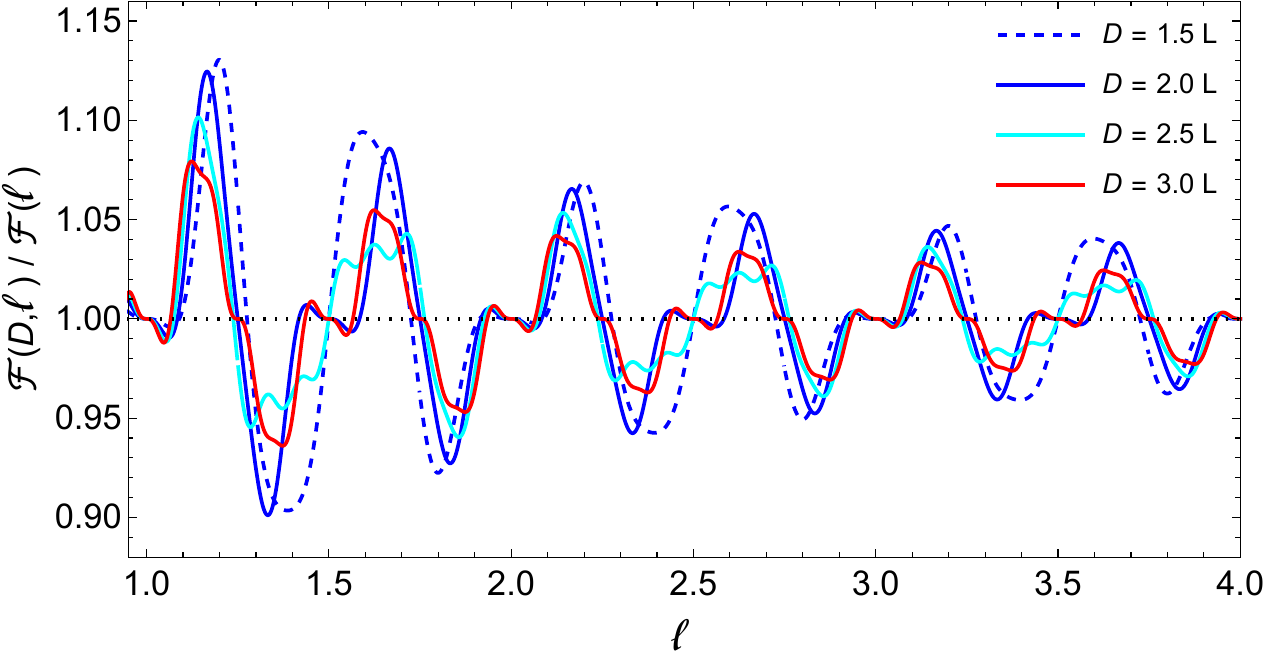}
    \caption{Ratio between the major shoaling length corrections in a quasi-linear approximation for the average value of the super-harmonic, one where the full integration interval $[-D,D]$ is taken into account and one restricted to the shoaling zone $[-L,0]$.}
    \label{Fig:elldx}
\end{figure*}
 \begin{eqnarray}
 \mathcal{H}_{11}^{(\textrm{I})}   &  =  &   \frac{ \sin{(4\pi d)} - \sin{(4\pi \ell)} }{4\pi (d -\ell)} \quad \,\, ; \quad   \mathcal{H}_{12}^{(\textrm{I})} = \frac{ \sin{(8\pi d)} - \sin{(8\pi \ell)} }{8\pi (d -\ell)}  \quad ,
\\
\mathcal{H}_{11}^{(\textrm{II})}   &  =  &   \frac{  \sin{(4\pi \ell)} }{4\pi \ell} \qquad \qquad \qquad \,\, ; \quad   \mathcal{H}_{12}^{(\textrm{II})} = \frac{  \sin{(8\pi \ell)} }{8\pi \ell}  \qquad \qquad \quad \,\, ,
\\
\mathcal{H}_{11}^{(\textrm{III})}   &  =  &  \frac{  \sin{(4\pi d)} }{4\pi d} \qquad \qquad \quad \quad \,\, ; \quad   \mathcal{H}_{12}^{(\textrm{III})} = \frac{  \sin{(8\pi d)} }{8\pi d}  \qquad \qquad \quad  .
\end{eqnarray}
Accordingly, we find the leading-order $i$-th normalized variance over all three zones (I, II, III), in the same manner we reached eq.~(\ref{eq:energyVAR}):
\begin{eqnarray}
\frac{2}{a^2} \langle \eta^2  \rangle^{(i)}     =  \big( 1 + \mathcal{H}_{11}^{(i)} \big)  + \big( 1 + \mathcal{H}_{12}^{(i)} \big) \left(  \frac{\pi \varepsilon \mathfrak{S} }{4} \right)^{2} \Tilde{\chi}_1^{(i)} \quad \quad . 
\label{eq:varGENx1}
\end{eqnarray}
Likewise, the normalized energy following eqs.~(\ref{eq:energyFIM}-\ref{eq:Gammal}) becomes:
\begin{eqnarray}
\frac{2}{a^2} \mathscr{E}^{(i)}   =  \big( 1 + \mathcal{H}_{11}^{(i)} \big)  + \big( 1 + \mathcal{H}_{12}^{(i)} \big) \left(  \frac{\pi \varepsilon \mathfrak{S} }{4} \right)^{2} \Big( \frac{\Tilde{\chi}_1^{(i)} + \chi_1^{(i)}}{2}  \Big) \quad \quad . 
\label{eq:varGENx2}
\end{eqnarray}
Because region I is at or near deep waters $(\mu \gtrsim \pi)$ while region III is near or at shallow waters $(\mu \lesssim 1/2)$, we have that $\Tilde{\chi}_1^{(\textrm{III})} \gg \Tilde{\chi}_1^{(\textrm{I})}$. As a result, we have that $\Tilde{\chi}_1^{(\textrm{II})} \approx (\Tilde{\chi}_1^{(\textrm{I})} + \Tilde{\chi}_1^{(\textrm{III})})/2 \approx \Tilde{\chi}_1^{(\textrm{III})} / 2$ is a good approximation for the range of relative water depth $\mu$ of interest. Taking these approximations and eqs.~(\ref{eq:varGENx1}-\ref{eq:varGENx2}) in mind, we generalize eq.~(\ref{eq:Gammal}) and the resulting $\Gamma$ for an arbitrary $d$ is:
\begin{equation}
\Gamma_{d} \approx \frac{ 1+   \frac{\pi^{2} \varepsilon^{2} \mathfrak{S}^2  }{16}     \, \Tilde{\chi}_{1} \mathcal{F}(d , \ell) }{  1+   \frac{\pi^{2} \varepsilon^{2} \mathfrak{S}^2  }{32}   \, \left( \Tilde{\chi}_{1}  + \chi_{1} \right) \mathcal{F}(d , \ell)   } \quad , 
\quad \frac{\mathcal{F}(d , \ell)}{\mathcal{F}(\ell)} \approx   \frac{  \big( 1 + \mathcal{H}_{11}^{(\textrm{II})} \big)  \left[  1  +  2 \left( \frac{ 1 + \mathcal{H}_{12}^{(\textrm{III})} }{  1 + \mathcal{H}_{12}^{(\textrm{II})}  } \right) \right] }{    3 + \mathcal{H}_{11}^{(\textrm{II})}   \left[ 1 + \frac{ \mathcal{H}_{11}^{(\textrm{I})} }{ \mathcal{H}_{11}^{(\textrm{II})} } + \frac{ \mathcal{H}_{11}^{(\textrm{III})} }{ \mathcal{H}_{11}^{(\textrm{II})} } \right]   } \quad ,
\label{eq:gammaLLx}
\end{equation}
where the correction due to the shoaling length $\mathcal{F}(d , \ell)$ is but oscillations around the base $\mathcal{F}(\ell)$ of eq.~(\ref{eq:Gammal}). As illustrated in \jfm{figure} \ref{Fig:elldx}, oscillations due to $d/\ell$ are quite weak. The magnitude of $|\mathcal{F}(d , \ell)/\mathcal{F}(\ell) - 1 | \ll 1$ demonstrates that the interval $[-L,0]$ contains sufficient physical information for the energetic analysis. Note that if we use the fully nonlinear evolution of the super-harmonic, the mean of $\mathcal{F}(d , \ell)/\mathcal{F}(\ell)$ will only be shifted upwards by $1-2\%$ and will have the same oscillations of \jfm{figure} \ref{Fig:elldx}, thus having a negligible impact on $\Gamma_{\ell}$.

\setcounter{equation}{0}
\section{Approximation of the reflection coefficient of a partially submerged steep slope}\label{append:Refl_coef}
The reflection rate \x{over a plane slope beach} according to \citet{Battjes1974} is:
\begin{equation}
    K_R \approx 0.1 \xi^2 \approx \frac{(\nabla h)^2}{10\varepsilon} \quad , \quad \xi \leq 2.5 \quad .
\label{eq:Battjes}
\end{equation}
On the other hand, \citet{Miche1951} found through theoretical arguments that,
\begin{equation}
K_R = \sqrt{\frac{2\alpha}{\pi}} \cdot \frac{\sin^2{\alpha}}{\pi \varepsilon} \quad .
\end{equation}
Nonetheless, in the modern context of coastal wave processes upon which most theoretical and numerical models are functions of the bottom slope (thus a tangent function). Ergo, the sine function above is, to say the least, inadequate for a comparison with other models.
Through a Taylor expansion up to third-order in $\alpha$ we can approximate:
\begin{equation}
\cos\alpha \equiv \frac{\sin{\alpha}}{\tan{\alpha}} = \frac{\sin{\alpha}}{\nabla h} =  1 - \frac{\alpha^2}{2} + O(\alpha^4)\, .
\end{equation}
therefore, $\sin{\alpha} \approx \nabla h(1 - {\alpha^2}/{2})$, and we get 
\begin{equation}
K_R = \sqrt{\frac{2\alpha}{\pi}}  \left( 1 - \frac{\alpha^2}{2}  \right)^2   \frac{(\nabla h)^2}{\pi \varepsilon} \quad .
\end{equation}
By further approximating the inverse of the tangent for steep slopes,
\begin{figure}
\centering
   \includegraphics[scale=0.65]{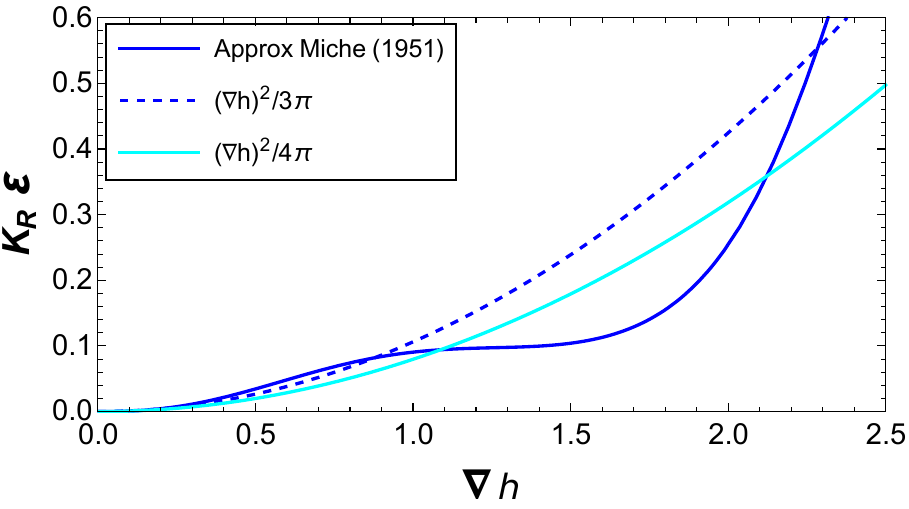}
   \caption{Consecutive approximations for \citeauthor{Miche1951}'s reflection coefficient.}
   \label{fig:Miche}
\end{figure}
\begin{equation}
\alpha = \tan^{-1}{\nabla h} \approx \nabla h \left[ 1 - \frac{(\nabla h)^2}{8}  \right] \,\, ,
\end{equation}
ergo,
\begin{equation}
\hspace{-0.3cm}
K_R  = \sqrt{\frac{2\nabla h}{\pi} \left[ 1 - \frac{(\nabla h)^2}{8}  \right]}  \left\{ 1 - \frac{(\nabla h)^2}{2} \left[ 1 - \frac{(\nabla h)^2}{8}  \right]^2  \right\}^2 \frac{(\nabla h)^2}{\pi \varepsilon }     \quad .
\end{equation}
Thus we may further approximate it and find to leading order (see \jfm{figure} \ref{fig:Miche}),
\begin{equation}
K_R = \sqrt{\frac{2\nabla h}{\pi}}  \left[ 1 - (\nabla h)^2 + \frac{(\nabla h)^4}{3} \right]   \frac{(\nabla h)^2}{\pi \varepsilon} \,\,    \sim   \,\,   \frac{(\nabla h)^2}{4 \pi \varepsilon}  \quad .
\label{eq:Miche}
\end{equation}
The estimation based on \citet{Miche1951} is valid only for mild slopes. The denominator is rather small because typically $\varepsilon \sim 1/20$ before breaking before a shoal, and thus, $4\pi \varepsilon \sim 1/2$. Accordingly, the reflection rate reaches 100\% too early even at reasonably steep slopes $\nabla h \geq 1/\sqrt{2}$ (equivalent of 35$^\circ$), and thus significantly overpredicts reflection rates \citep{Zanuttigh2008}. Other theoretical models have found the slope-dependent reflection coefficient to be described as an exponential of the type $e^{-1/(\nabla h)^2}$ \citep{Miles1990,Ehrenmark1998}, which could in principle still be approximated as quadratic. Although several methods exist to compute the reflection rate, especially for the case of a step instead of a finite slope, \citet{Chang2004} has shown that a comparison between the most popular methods still finds all of them to be compatible with the term $(\nabla h)^2$.

\begin{figure*}
\minipage{0.4\textwidth}
    \includegraphics[scale=0.65]{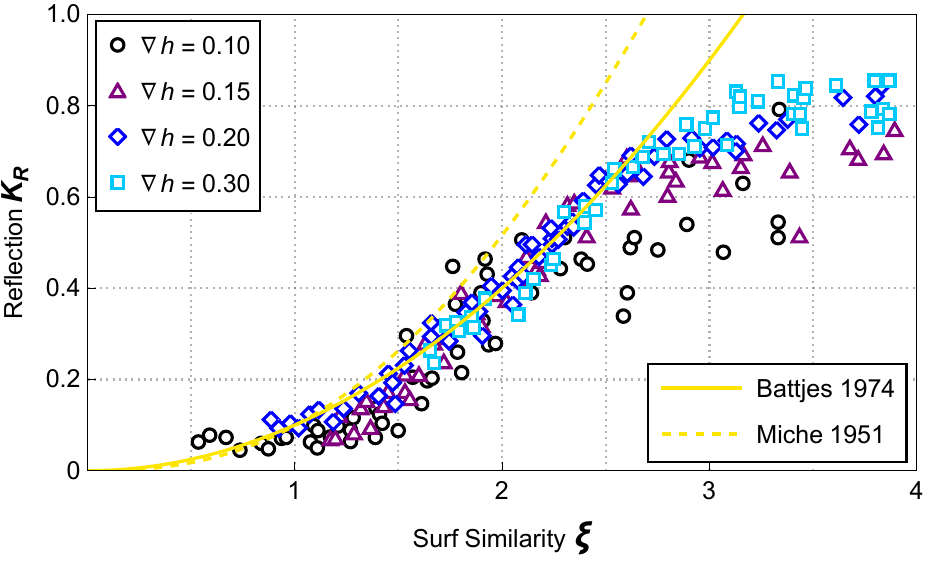}
\endminipage
\hfill
\minipage{0.4\textwidth}
    \includegraphics[scale=0.7]{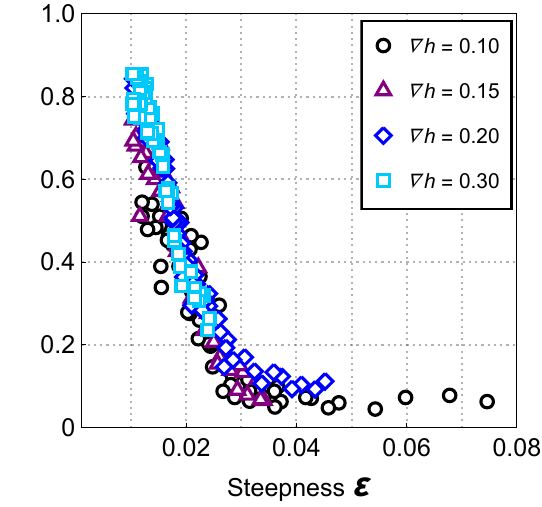}
\endminipage
   \caption{(a) Rendering of Fig 2.8 of \citet{Battjes1974} comparing its own formula to the data collected in \citet{moraes1970experiments}. (b) Inversion of the data to extract wave steepness.}
\label{fig:Surf1}   
\end{figure*}

From the perspective of \citet{Battjes1974}, the overprediction by reflection models is related to their range of validity. Indeed, \jfm{figure}~\ref{fig:Surf1}\jfm{a} indicates that the reflection coefficient of eq.~(\ref{eq:Battjes}) performs quite well for all experiments with different slopes $\nabla h \in [0.1,0.3]$ as long as $\xi <2.5$. Otherwise, \citeauthor{Battjes1974}'s analytical model substantially overpredicts the reflection rate. As a note, the data can not be fitted with a linear function because in that case the initial value of reflection at $\xi \rightarrow 0$ has to be negative. \jfm{Figure}~\ref{fig:Surf1}\jfm{b} provides supplementary information: inversion of the surf similarity equation implies that the data from \citet{moraes1970experiments} depicts low wave steepness behind the high values of $\xi$ being overpredicted, whilst moderate to high wave steepness is associated to the regime in which \citeauthor{Battjes1974}'s model is in excellent agreement with observation.  Thus, we conclude that when the initial steepness is far from a wave-breaking condition, we find that for very steep slopes $\nabla h \sim 1$ one has $\xi > 3$ and the reflection rate formula has to be corrected. 

\begin{figure*}
\centering
   \includegraphics[scale=0.64]{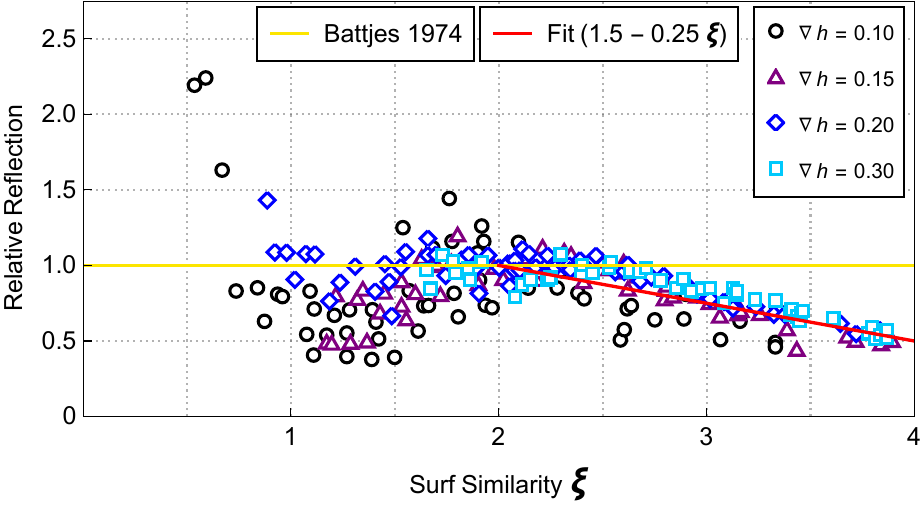}
   \caption{Adaptation of \jfm{figure}~\ref{fig:Surf1} to compute the ratio between observed reflection rates to the theoretical ones of \citet{Battjes1974}. Solid red line shows the downward trend of the ratio for high surf similarity.}
   \label{fig:Surf2} 
\end{figure*}
\begin{table*}
\centering
\begin{tabular}{cccccccccccc}    
\toprule
\emph{$\frac{2\sigma}{\langle K_{R} \rangle}$} &   \quad \emph{$0.58$} \quad &  \quad \emph{$0.53$} \quad &  \quad \emph{$0.48$} \quad & \quad \emph{$0.30$} \quad & \quad \emph{$0.27$} \quad & \quad \emph{$0.23$} \quad & \quad \emph{$0.12$} \quad 
\\
\midrule
$\langle \xi \rangle$  & \quad 0.81 \quad  & \quad 1.28 \quad & \quad 1.68 \quad  & \quad 2.24 \quad & \quad 2.73 \quad & \quad 3.24 \quad & \quad 3.78 \quad \\
$\sigma$  & \quad 0.02 \quad  & \quad 0.04 \quad & \quad 0.06 \quad  & \quad 0.08 \quad & \quad 0.10 \quad & \quad 0.11 \quad & \quad 0.06 \quad
\\
\bottomrule
\end{tabular}
\caption{Variation on two standard deviations of \citeauthor{moraes1970experiments}'s data compared to the mean reflection coefficient for subsets of surf similarity data.} 
\label{tab:Ksigma}
\end{table*}

Perhaps an even more revealing analysis of the aforementioned data is presented in \jfm{figure}~\ref{fig:Surf2}: we compute the ratio between observed and analytical (modelling) reflection rates as a function of the surf similarity. In the interval $1.5 < \xi < 2.5$ the agreement between model and observation is ideal and belongs to a reasonable uncertainty level $0.8 < K_{R \, , obs} / K_{R}< 1.2$. On the contrary, for $\xi < 1.5$ we see some strong scatter. However, in the latter regime the absolute differences are very small because the reflection rates are tiny. Whether the reflection rate is 5\% or 15\% has no impact on the
wave statistics. The crucial regime is hence $\xi \geqslant 2.5$ because the difference in reflection rates is high (20-30\% difference between observation and model), and the distinction between a reflection rate of 20\% and 50\% is substantial for wave statistics \citep{Yuchen2023}. For the purposes of measuring how much the data scatters, using data of \jfm{table}~\ref{tab:Ksigma} for the evolution of standard deviation on reflection rates, we can estimate a 95\% confidence interval on the latter ($\sigma$ the standard deviation),
\begin{equation}
\frac{2\sigma}{K_{R \, , obs}} \approx  \frac{5- \xi}{7} \quad , \quad \xi \leqslant 4 \quad .
\label{eq:Kstd}
\end{equation}
Additionally, \jfm{figure}~\ref{fig:Surf2} features a worrisome trend (red solid line) in which overprediction of reflection rates grows linearly wth $\xi$ regardless of the slope magnitude. The experimental trend can be approximated by,
\begin{equation}
\frac{K_{R \, , obs}}{ K_{R}} \approx \frac{6 - \xi}{4}  \quad , \quad \xi \geqslant 2 \quad .
\label{eq:Ktrend}
\end{equation}
It is then of interest to find a number or function $\mathfrak{B}$ more suitable for very steep slopes,
\begin{equation}
K_R = \mathfrak{B} \cdot (\nabla h)^2 \equiv \tilde{\xi}^2 = \frac{\tilde{\mathfrak{B}}}{\varepsilon} (\nabla h)^2 \quad .
\label{eq:NEWreflec}
\end{equation}
This new coefficient generalizes the special case $\mathfrak{\tilde{\mathfrak{B}}}^{-1} = 10$ that recovers \citeauthor{Battjes1974}'s model. For instance, the results of \citet{Seelig1981} imply that we can estimate an upper bound for the reflection rate of waves traveling past very steep bottom slopes and mild wave steepness compatible with simulation cases described in \jfm{table} \ref{tab:cases3},
\begin{equation}
     K_R \lesssim \frac{1}{3} (\nabla h)^2 \quad , \quad  \mathfrak{B} \lesssim \frac{1}{60\varepsilon} \quad . 
\end{equation}
Combining both sets of experimental evidence, we can find proper upper and lower bounds using the trend encoded in eq.~(\ref{eq:Ktrend}). As we are interested in making bottom slopes steeper than simulated in the runs of \jfm{table}~\ref{tab:cases3} whilst maintaining the wave steepness values, we may use $\nabla h^{\dagger} = n \cdot \nabla h$ to denote steeper slopes than the reference values $1/20 < \nabla h < 1/3$ in the literature \citep{moraes1970experiments,Battjes1974,Seelig1981}. Wherefore, the accuracy of the model will depend on the pair $(n , \mathfrak{B})$:
\begin{equation}
\frac{K_{R \, , obs}}{ K_{R}} \approx \frac{(6 - n \tilde{\xi})}{4}  \quad , \quad \xi \gg 1 \quad .
\label{eq:Ktrend2}
\end{equation}
Note that the surf parameter within brackets in eq.~(\ref{eq:Ktrend2}) has to be much smaller than $\xi$, otherwise the maximum possible surf similarity becomes $\xi_{c} = 6/n$, suggesting the slope can not be twice steeper than in \citet{Battjes1974}. Thus, we formulate our problem as finding a value for $\tilde{\mathfrak{B}}$ such that for any slope we have observation and model converging:
\begin{figure*}
\centering
   \includegraphics[scale=0.45]{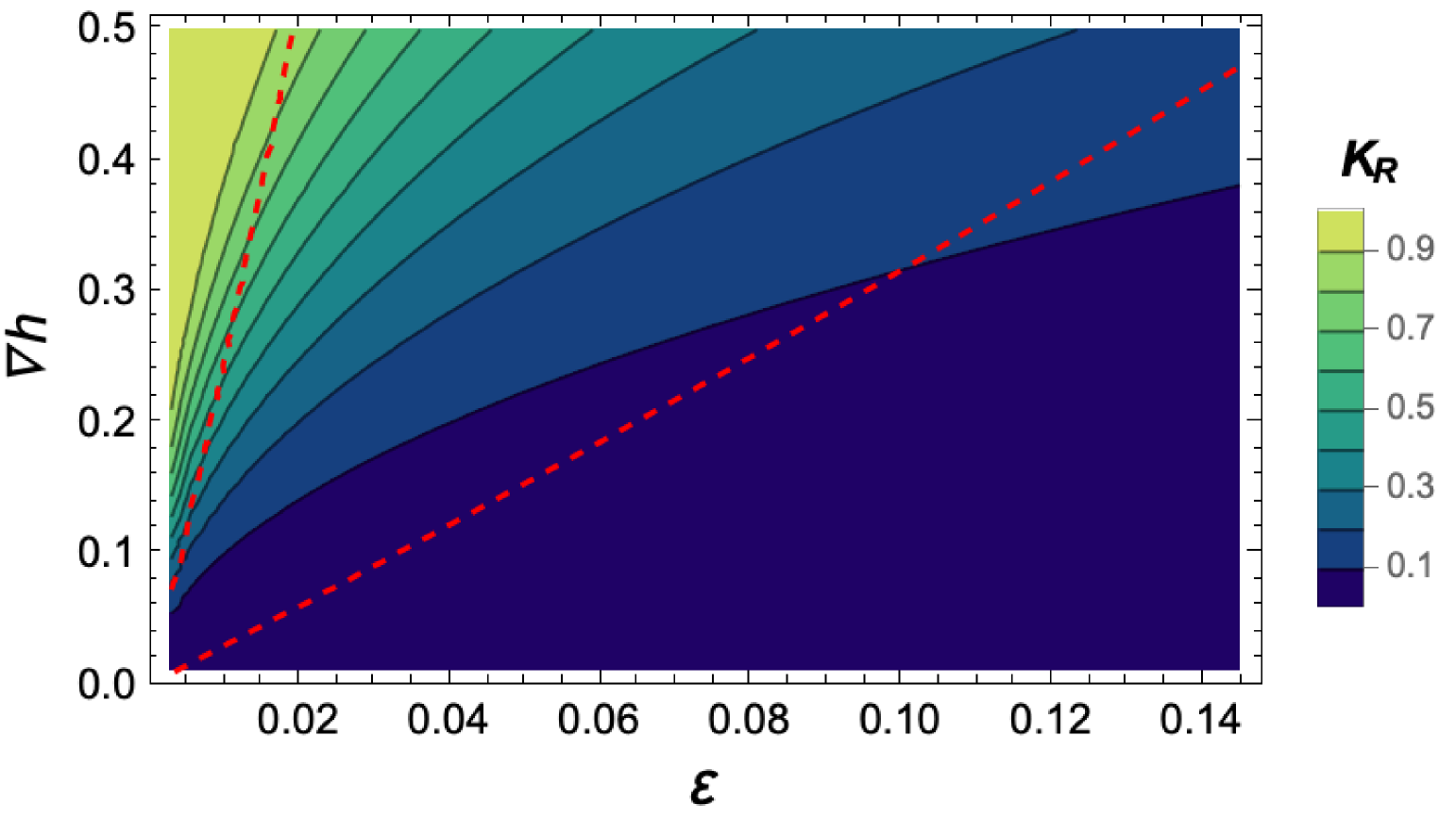}
   \caption{Contour plot of reflection $K_{R}$ based on the model of \citet{Seelig1981} compared to the contours of eq.~(\ref{eq:OurMODEL}) for two values $K_{R} = (0.1,0.8)$ in red dashed lines.}
   \label{fig:Miche5}
\end{figure*}
\begin{equation}
\lim_{\xi \rightarrow \infty} K_{R \, , obs} = \lim_{n \rightarrow \infty} K_{R \, , obs} = 1 
\end{equation}
\citet{Seelig1981} addresses this problem with a numerically based generalization of eq.~(\ref{eq:Battjes}) through a hyperbolic tangent function, thus recovering the well-behaved model at $\xi < 2.5$ and converging the reflection rate to unity at large surf similarities $\xi \gg 1$. However, \citeauthor{Seelig1981}'s purpose was the reflection rate in itself, whereas here we are interested in a model that is useful for wave statistics. For the latter, a polynomial approximation is much preferred. Taking eq.~(\ref{eq:NEWreflec}) into account, we rewrite eq.~(\ref{eq:Ktrend2}) as follows:

\begin{equation}
K_{R \, , obs} = \frac{\tilde{\mathfrak{B}}}{\varepsilon} (\nabla h)^2  \cdot \frac{1}{4} \Bigg[ 6 - n \sqrt{\tilde{\mathfrak{B}}} \xi  \Bigg]
\end{equation}
To avoid obtaining a function such that $\tilde{\mathfrak{B}} (\nabla h)^2$ is a constant, we use the reference surf similarity where disparities between theory and observation arise and the order of magnitude of the bottom slope we want to reach $\nabla h^{\dagger} \sim 3 \nabla h \sim 1$ for the range in steepness similar to entries in \jfm{table}~\ref{tab:cases3}, such that we simplify the problem:
\begin{equation}
\lim_{n \rightarrow \infty} K_{R \, , obs} = 5\tilde{\mathfrak{B}} \Big( 6 - 3 n \sqrt{\tilde{\mathfrak{B}}}   \Big) \approx 1 \quad .
\end{equation}
The above relation will lead to a cubic polynomial equation, which we shall avoid. Rather, considering that $\tilde{\mathfrak{B}} \ll 1$, we shall approximate $\sqrt{\tilde{\mathfrak{B}}} \approx 3\tilde{\mathfrak{B}}$. Then, the solution after a few iterative approximations that recovers the initial condition $\tilde{\mathfrak{B}} (n=1) \approx 0.1$ is best written as,
\begin{equation}
\tilde{\mathfrak{B}} \approx \frac{5+ \sqrt{25-5n}  }{90n} \approx \frac{1+ \sqrt{1-\frac{n}{5}}  }{18n} \approx \frac{20-n}{180n} \quad .
\end{equation}
Further considering that compared to the reference slopes of \jfm{figure}~\ref{fig:Surf1} we have that $n \approx 10\nabla h / 3$, we thus finally obtain a polynomial approximation that extends \citeauthor{Battjes1974}'s model for the reflection rate to very steep slopes:
\begin{equation}
K_R (\nabla h) := \frac{\tilde{\mathfrak{B}}}{\varepsilon} (\nabla h)^2 \approx  \left( \frac{6}{\nabla h} - 1 \right) \frac{(\nabla h)^2}{180\varepsilon}   \quad , \quad \xi \geqslant 2.5 \quad .
\label{eq:OurMODEL0}
\end{equation}
Considering the typical wave steepness $\varepsilon \sim 1/40$ from \jfm{figure}~\ref{fig:Surf1}\jfm{b}, we can compute reflection rates from all the models here presented in addition \citep{Miche1951,Battjes1974} to that of \citet{Seelig1981}, the latter reading $K_{R} = \tanh{(0.1 \xi^2)}$. In \jfm{figure}~\ref{fig:Miche4} we replot \citeauthor{moraes1970experiments}'s data against the four models, and we observe that \citeauthor{Miche1951}'s is accurate up to $\xi <2$, \citeauthor{Seelig1981}'s model certainly improves that of \citet{Battjes1974} and extends its accuracy up to $\xi < 4$. Interestingly, our semi-analytical model performs closely to that of \citet{Seelig1981}, providing slightly lower reflection rates than the latter for high values of the surf similarity. 
The new model of eq.~(\ref{eq:OurMODEL}) can be readily transformed into a surf similarity one through eq.~(\ref{eq:upxi2}):
\begin{equation}
K_R (\xi)  \approx  \left( \frac{6 }{\xi \sqrt{\varepsilon}} - 1 \right) \frac{\xi^2}{180}  \quad .
\label{eq:OurMODEL}
\end{equation}
Remarkably, although eq.~(\ref{eq:OurMODEL}) is expected to be valid only for $\xi > 2.5$, it actually performs reasonably well across the entire range. Whenever it overpredicts reflection rates for low surf similarity, the rates are in absolute values small. This somewhat small deviation bears little impact for wave statistics as it belongs to the regime of eq.~(\ref{eq:refKurt}). Withal, it behooves us to model either upper bound or lower bound of the shaded area in \jfm{figure}~\ref{fig:Miche4} with a simpler quadratic function of either the bottom slope or surf similarity. In fact, \citeauthor{Seelig1981}'s formula has no variable coefficient as functions of $(\xi, \nabla h)$. Thus, a lower bound on the reflection can be obtained with a quadratic polynomial because the data already behaves thusly, and all one needs to find is the smallest coefficient valid for $\xi \gg 1$ that also reasonably describes the physics at $\xi \leqslant 1$. However, an upper bound with a purely quadratic function is not feasible because of the concavity of the shaded upper bound in \jfm{figure}~\ref{fig:Miche4}. To do so, we seek a reference bounded value for $6/\xi_{\infty}\sqrt{\varepsilon}$ such that the entire formula matches the lower bound. For the experiments of \citet{moraes1970experiments}, this value is somewhere near $\xi_{\infty} = 4$, and as such we have,
\begin{equation}
K_{R}^{(-)}  \approx  \left( \frac{6 }{\xi_{\infty} \sqrt{\varepsilon}} - 1 \right) \frac{\xi^2}{180} =  \left( \frac{3 }{2 \sqrt{\varepsilon}} - 1 \right) \frac{\xi^2}{180}   \quad .
\label{eq:OurMODEL2}
\end{equation}
\jfm{Figure}~\ref{fig:Miche5} reinforces this assessment by comparing the reflection rates of \citeauthor{Seelig1981}'s and eq.~(\ref{eq:OurMODEL}) in a contour plot for any wave steepness and bottom slope. It becomes apparent that it is possible to use polynomial model for the reflection rate that performs nearly as well as the hyperbolic function of \citet{Seelig1981}.

\bibliography{Maintext}

\end{document}